\documentclass[journal]{IEEEtran}

\usepackage{amsmath}
\usepackage{graphicx}

\usepackage{flushend}
 
\usepackage{latexsym}
\usepackage{amsfonts}
\usepackage{graphicx}   
\usepackage{subfigure}
\usepackage{multirow} 
\usepackage{makecell}
\usepackage{amssymb} 
\usepackage{enumerate}
\usepackage{amsmath}
\usepackage[usenames]{color}
\usepackage[dvipsnames]{xcolor}
\usepackage{longtable}
\usepackage{tabularx}
\usepackage{chngpage}
\usepackage{lipsum}  
\usepackage{booktabs}
\usepackage{subfigure} 
\usepackage{verbatim}
\usepackage{xspace}

\usepackage{tikz}
\usetikzlibrary{arrows}

\usepackage{bm}
\usepackage{pdflscape}

\newcommand{\Real}{\mathbb{R}}

\newcommand{\bSigma}{\bm{\Sigma}}

\newcommand{\bmu}{\bm{\mu}}
\newcommand{\bnu}{\bm \nu}

\newcommand{\bb}{{\bf b}}
\newcommand{\x}{{\bf x}}
\newcommand{\y}{{\bf y}}

\newcommand{\X}{{\bf X}}
\newcommand{\bC}{{\bf C}}
\newcommand{\bI}{{\bf I}}
\newcommand{\bA}{{\bf A}}

\newcommand{\cblue}{\textcolor{black}}

\newcommand{\normalized}{\widetilde}

\newcommand{\E}{\mathrm{E}}

\newcommand\acro{O-PMC}
\def\x{{\mathbf x}}

\begin{document}

\title{Optimized Population Monte Carlo}

\author{V\'ictor~Elvira,~\IEEEmembership{Senior~Member,~IEEE,}
        and~\'Emilie Chouzenoux,~\IEEEmembership{Senior~Member,~IEEE}
\thanks{V. Elvira is with the School of Mathematics at the University of Edinburgh (UK) and with The Alan Turing Institute (UK). \'E Chouzenoux is with Universit\'e Paris-Saclay, Inria, CentraleSup\'elec, Centre de Vision Num\'erique (France).}% <-this % stops a space
\thanks{V.E. and \'E.C. acknowledge support from the \emph{Agence Nationale de la Recherche} of France under PISCES (ANR-17-CE40-0031-01) and MAJIC (ANR-17-CE40-0004-01) projects. V.E. acknowledges support from ARL/ARO under grant W911NF-20-1-0126. \'E.C. acknowledges support from the European
Research Council Starting Grant MAJORIS ERC-2019-STG-850925.}% <-this % stops a space
}

%\markboth{IEEE TRANSACTIONS ON SIGNAL PROCESSING,~Vol.~X, No.~Y, DATE}%
{ }
\markboth{}%
{ }

\maketitle

\begin{abstract}
Adaptive importance sampling (AIS) methods are increasingly used for the approximation of distributions and related intractable integrals in the context of Bayesian inference. Population Monte Carlo (PMC) algorithms are a subclass of AIS methods, widely used due to their ease in the adaptation. In this paper, we propose a novel algorithm that exploits the benefits of the PMC framework and includes more efficient adaptive mechanisms, exploiting geometric information of the target distribution. In particular, the novel algorithm adapts the location and scale parameters of a set of importance densities (proposals). At each iteration, the location parameters are adapted by combining a versatile resampling strategy (i.e., using the information of previous weighted samples) with an advanced optimization-based scheme. Local second-order information of the target distribution is incorporated through a preconditioning matrix acting as a scaling metric onto a gradient direction. A damped Newton approach is adopted to ensure robustness of the scheme. The resulting metric is also used to update the scale parameters of the proposals. We discuss several key theoretical foundations for the proposed approach. Finally, we show the successful performance of the proposed method in three numerical examples, involving challenging distributions.
\end{abstract}

\begin{IEEEkeywords}
Importance sampling, Monte Carlo methods, population Monte Carlo, Newton algorithm, covariance adaptation, stochastic optimization, Langevin dynamics.
\end{IEEEkeywords}

%%%%%%%%%%%%%%
\section{Introduction}
\label{sec_intro}
%%%%%%%%%%%%%%
%\input{intro.tex}

Intractable integrals appear in countless problems of science and engineering. For instance, in Bayesian inference the interest is in estimating a posterior distribution of an unknown parameter given a set of related data. For most realistic models, the posterior distribution cannot be obtained in a closed form, and even more, it is not possible to simulate samples from it. 
Therefore, obtaining moments of interests (e.g., the mean, the variance, the probability of a certain event) is unfeasible either via an exact closed form or through approximations involving direct sampling.  
Importance sampling (IS) is a popular type of Monte Carlo methods \cite{Robert04,Liu04b,Owen13} for the approximation of intractable distributions and related integrals. 
In its standard procedure, IS requires the simulation of samples from another distribution (called proposal). The samples receive an importance weight that takes into account the mismatch between target and proposal distributions.
IS is a theoretically solid mechanism with strong guarantees, such as consistency, central limit theorems, and explicit error bounds \cite{Owen13,elvira2021advances}.  
The performance in IS strongly depends on the adequacy of the proposal distribution. 
Intuitively, a proposal is good when it is \emph{close} to the integrand in the targeted integral.
However, it is usually impossible to know in advance where the probability mass of the target distribution is located (e.g., in Bayesian inference, one only has access to the evaluation of an unnormalized version of the posterior distribution).
Therefore, advanced strategies must be employed, usually involving more than one proposal, which is called multiple importance sampling (MIS) \cite{elvira2019generalized}, combined with the adaptation of the multiple proposals, leading to adaptive importance sampling (AIS)  schemes \cite{bugallo2017adaptive}. %\emilie{please check if this reformulation is correct.}

The literature of AIS is vast, including methods based on sequential moment matching such as AMIS \cite{CORNUET12,marin2019consistency}, that comprises a Rao-Blackwellization of the temporal estimators, and APIS that incorporates multiple proposals \cite{APIS15}. Other recent methods have introduced Markov chain Monte Carlo (MCMC) mechanisms for the adaptation of the IS proposals \cite{martino2017layered,schuster2018markov,rudolf2020metropolis}. 
The family of population Monte Carlo (PMC) methods also falls within AIS. Its key feature is arguably the use of resampling steps in the adaptation of the location parameters of the proposals \cite{Douc05,li2015resampling}.
The seminal paper \cite{Cappe04} introduced the PMC framework. Since then, other PMC algorithms have been proposed, increasing the resulting performance  by the incorporation of stochastic expectation-maximization mechanisms \cite{Cappe08}, non-linear transformation of the importance weights \cite{koblents2013robust}, or better weighting and resampling schemes \cite{elvira2017improving}. 
%
%One of the main limitations of these methods is that they rarely adapt the scale parameters.  
%
The method we propose in this paper falls within the PMC framework.
%
%\emilie{Improve transition here. We jump from PMC to AIS. What was the motivation for location parameter adaptation in AIS ?}

The state-of-the-art AIS methods, and particularly those belonging to the PMC family, suffer from several limitations that prevent a wider application of IS to more challenging problems. 
First, in the case of PMC, the resampling step provokes the well-known path degeneracy (see for instance \cite[Fig. 4]{elvira2017improving}), endangering the diversity of the proposals in the subsequent iterations. 
Some attempts have been recently done to attenuate this problem, e.g., the LR-PMC in \cite{elvira2017improving} first forms a partition of the samples and then performs independent resampling step in each subset. However, this is at the expense of  worsening the local exploration, since each partition approximates the target with less samples (see more details in \cite{elvira2017improving}). 
The second limitation, not only in PMC but also in AIS in general, is that most existing methods only adapt the location parameters {of the proposals}, while the scale parameter remains fixed from the beginning.  %\emilie{First time we talk about scale parameter. Is it clear enough it related to proposal ? Maybe insert (earlier) a simple statement to say a proposal depends on scale and location parameters ?}. %Second, most of PMC methods only adapt the location parameter but are unable to fit the scale parameter.
This is a clear limitation, since {it is well known that the scale parameters of the proposals can make a significant difference in the efficiency of the AIS algorithm. Moreover a fine manual tuning requires a prior knowledge about the scale of the posterior distribution. Finally, even if such optimal fine tuning was possible, there is a clear advantage in adapting the scale parameter over the iterations, depending where the proposals are placed.} % and a manual fine tuning. 
Moreover, this represents an extra challenge when the dimensions of the posterior are of different order of magnitude and/or present strong correlations.
%
%\victor{From here it is repetitive. I would move it down and adapt the second and third sentences, to talk about our method. REMOVED 19jan}
%It is interesting to note that most works focus on either only location or also location and scale parameters adaptation. The latter case allows for a fully adaptation of Gaussian proposals, which are considered in most existing works. It is also possible to adapt other families of proposals by fixing the extra parameters (e.g., the degrees of freedom in Student's $t$-distribution) while adapting location and scale parameters.

{Some families of AIS methods use geometric information about the target for the adaptation of the location parameters, yielding to optimization-based adaptation schemes.} 
For example, the GAPIS algorithm \cite{elvira2015gradient} is an AIS method that exploits the gradient and the Hessian of the logarithm of the target, and also introduces an artificial repulsion among proposals to promote the diversity (without any resampling step). Other methods such as \cite{schuster2015gradient,fasiolo2018langevin} adapt the location parameters by performing at each sample several steps of the unadjusted Langevin algorithm (ULA) \cite{Roberts_MALA}, which can also be seen as an instance of a stochastic gradient descent method. The covariance is also adapted in those methods by either computing the sample autocorrelation \cite{schuster2015gradient} or using second-order information \cite{elvira2015gradient,fasiolo2018langevin}. A covariance adaptation has been also explored via robust moment-matching mechanisms in \cite{el2018robust,el2019recursive}. We refer the interested reader to the survey \cite{bugallo2017adaptive}. {The use of optimization techniques within PMC framework remains however unexplored.}
%\emilie{reading this, it seems Fasiolo et al do the same than us, in terms of proposal adaptation. We should double check the differences.} \
%
It is worth mentioning that optimization inspired schemes have also shown to be an efficient strategy to improve practical convergence rate in MCMC algorithms (see the survey paper \cite{Pereyra16} and references therein). In particular, the works \cite{roberts2002langevin,durmus2016efficient,Schreck16,qi2002hessian,Fisher_MCMC_2} fall in the framework of the so-called Metropolis adjusted Langevin algorithms (MALA), where the ULA scheme is combined with a Metropolis-Hastings step. The Langevin-based strategy yields proposed samples that are more likely drawn from a highly probable region, with the consequence of a larger acceptance probability. MALA can be further improved by rescaling the drift term by a preconditioning matrix encoding local curvature information about the target density, through the Fisher metric \cite{Fisher_MCMC_1}, the Hessian matrix \cite{Newton_MCMC_1,Newton_MCMC_2,qi2002hessian} or a tractable approximation of it~\cite{marnissi2018,Marnissi2014Eusipco,MarnissiEntropy2018,Simsekli2016}. Optimization-based methods for accelerating MCMC sampling of non-differentiable targets have also been considered, for instance in \cite{durmus2016efficient,Vono2019}. 

{
 In this work, we propose a new Optimized PMC (\acro) approach.\footnote{A limited version of this work was presented by the authors in the conference paper \cite{ElviraIcassp2019}.} {To the best of our knowledge, the proposed algorithm is the first within the relevant PMC family to incorporate explicit optimization steps in order to enhance the resampling-based adaptation by exploiting the geometry of the target. In \acro,} the proposals are adjusted using a {stochastic Newton-based step} adapted to the sample values resulting from a {suitable resampling} strategy. In contrast to the aforementioned works, here the mean and covariance adaptation are performed jointly, with the advantage of fitting the proposal distributions locally, boosting the exploration and increasing the performance. A damped Newton strategy, {incorporating two stabilization features} is proposed for the mean adaptation, and the retained scale matrix is per-used for the covariance adaptation. We show on three sets of numerical examples that this novel methodology {catalyzes} the local adaptation without endangering the diversity of the proposals {nor the stability of the trajectories}. % \emilie{I want to say that it cannot numerically diverge, as GAPIS could, for eg}.  

{
The rest of the paper is structured as follows. Section~2 introduces the problem setting, the AIS framework{, and optimization-based proposal adaptation rules}. In Section 3, we present the proposed method. We discuss its rationale and theoretical foundations in Section 4, {including also a toy example}. Finally, we show three numerical examples in Section 5 and conclude in Section~6.} 

%\victor{what should we do with this list?
%Main features:
%\begin{itemize}
	%\item $K$ samples per proposal (already in 2017)
	%\item alternative resampling schemes (local (already in 2017) or hybrid or something new)
	%\item DM weights (already in 2017)
	%\item mean and *covariance* Newton 
%\end{itemize}
%\cgreen{Insist more about the novelties.}
 %}
%%%%%%%%%%%%%%%%%%%%%%%%
\section{Bayesian inference via importance sampling}
\label{sec_bayesian}

In this section, we describe the Bayesian inference framework, the generic importance sampling methodology, and the standard PMC, which is an adaptive IS (AIS) algorithm. Note that, as stated in the introduction, the range of applicability of \acro~goes beyond Bayesian inference (e.g., in the first two examples presented in Section \ref{sec_exp}, the target distribution is available in a closed form and not necessarily coming from a Bayesian inference problem). %\emilie{maybe our simulations should illustrate this ? Or give a ref where PMC methodology is used not for Bayesian inference ?} \victor{This is for any Monte Carlo method, the involved distribution can be anything. Our simulations already show that.}
 
\label{sec:problem}
%%%%%%%%%%%%%%%%%%%%%%%%%
\subsection{Bayesian inference}

We consider the estimation problem of a vector of unknowns $\x\in\Real^{d_x}$ that is statistically connected through a probabilistic model to the vector $\y\in\Real^{d_y}$ that contains the available data. The observation model is embedded into the likelihood function $\ell(\y|\x)$. The Bayesian approach allows for the incorporation of available prior information about $\x$ in the so-called prior distribution $p_0(\x)$. The so-called posterior distribution of the unknowns given the data (a.k.a. \emph{target} distribution) can then be expressed thanks to the Bayes rule:
\begin{equation}
	\normalized{\pi}(\x| \y)
		= \frac{\ell(\y|\x) p_0(\x)}{Z(\y)}. %=\ell(\y|\x) p_0(\x),
\label{eq_posterior}
\end{equation}
Very often, the interest lies in the computation of a specific moment of the posterior distribution which amounts to solving integrals under the generic form
\begin{equation}
	I =  \int h(\x) \normalized{\pi}(\x) d\x = \frac{1}{Z} \int h(\x) \pi(\x) d\x,
\label{eq_integral}
\end{equation}
where $h$ is any integrable function w.r.t. $\widetilde \pi(\x)$. 
Unfortunately, in most cases of interest, Eq. \eqref{eq_integral} cannot be computed, either because the integral is intractable or because the posterior distribution is rarely available in a closed form, mostly because of the impossibility of computing the normalizing constant $Z(y) \triangleq \int  \pi(\x|\y) d\x$ (a.k.a. model evidence, marginal likelihood, or partition function). Hence, it is useful to define the non-negative function $\pi(\x|\y)\triangleq\ell(\y|\x) p_0(\x) =  Z(\y) \normalized\pi(\x| \y)$. \cblue{From now on, in order to ease the notation, we drop $\y$ in $Z$, $\pi(\x)$, and $\widetilde \pi(\x)$.}  %\emilie{I am lost in those notations. Can you check if there is not a tilde missing in Z(y) definition ?}
In order to overcome this limitation, approximate methods must be employed.

\subsection{Importance sampling}
\label{sec_IS}
Importance sampling (IS) is a Monte Carlo methodology that allows for the approximation of distributions and integrals as those of previous section. Unlike the raw (or standard) Monte Carlo technique, the basic IS method simulates all samples from the so-called \emph{proposal} distribution $q(\x)$. The samples are weighted accordingly in such a way consistent estimators can be built. More precisely, IS is composed of the two following steps:

\begin{enumerate}
	\item \textbf{Sampling.} Simulate $K$ samples as 
	$$\x_k \sim q(\x), \qquad k=1,...,K.$$ % add i.i.d.
	\item \textbf{Weighting.} Assign an importance weight to each sample as
	$$w_k = \frac{\pi(\x_k)}{q(\x_k)} \qquad k=1,...,K.$$
	i.e., the ratio between the unnormalized target and the proposal distribution, evaluated at the specific sample.
\end{enumerate}
This basic sampling-weighting procedure allows for the construction of the both next estimators:
\begin{itemize}
	\item Unnormalized IS (UIS) estimator:
	\begin{equation}
	\hat I = \frac{1}{KZ} \sum_{k=1}^{K} w_k h(\x_k).
	\label{eq_UIS}
	\end{equation}
	\item Self-normalized IS (SNIS) estimator:
\begin{equation}
\widetilde I = \sum_{k=1}^{K} \bar w_k h(\x_k).
\label{eq_SNIS}
\end{equation}
\end{itemize}
Both UIS and SNIS are consistent with $K$, while only UIS is unbiased. However, UIS can be used only if $Z$ is known.  The key of success in IS is an appropriate choice of the proposal $q$ in such a way that the aforementioned estimators have a low variance.  The variance of the UIS estimator is minimized when the proposal is $q(\x) \propto |h(\x)|\pi(\x)$, while the optimal proposal of the SNIS estimator is $q(\x) \propto |h(\x)|\pi(\x)$ \cite{Robert04,Liu04b,Owen13}. However, in most of cases it is impossible to design such proposal because it does not have a known form where sampling is possible. Hence, adaptive methods are required in order to iteratively improve the proposal.

\subsection{Multiple importance sampling}
\label{sec_mis}
Multiple importance sampling (MIS) refers to the case where several proposals $\{q_n(\x)\}_{n=1}^N$ are used instead of just one, as in the previous section. It is known that in MIS, many possible sampling and weighting schemes are possible, and we refer the interested reader to an exhaustive comparison and analysis in~\cite{elvira2019generalized}. Let us consider the case where $K=N$ samples are simulated from the set of $N$ proposals. One can proceed as follows:
\begin{enumerate}
\item \textbf{Sampling:} Each sample is simulated from each of the proposals as 
$$\x_n \sim q_n(\x),\qquad n=1,...,N$$
\item  \textbf{Weighting:} Among all possible weighting options, we describe two possibilities:
\begin{itemize}
\medskip
\item \textbf{Option 1}: Standard MIS (s-MIS): 

$$w_n= \frac{\pi({\bf x}_n)}{{q_n({\bf x}_n)}}, \quad n=1,\ldots, N $$% \label{weights_SIS}
\medskip
\item \textbf{Option 2}: Deterministic mixture MIS (DM-MIS):
\begin{equation} 
	w_n=\frac{\pi({\bf x}_n)}{\psi({\bf x}_n)}=\frac{\pi({\bf x}_n)}{\frac{1}{N}\sum_{j=1}^{N}q_j({\bf x}_n)}, \quad n=1,\ldots, N,\nonumber
\label{f_dm_weights_static}
\end{equation}
where $\psi({\bf x})=\frac{1}{N}\sum_{j=1}^{N}q_j({\bf x})$ is the mixture pdf. 
% \begin{equation} 
% 	w_n= \frac{\pi({\bf x}_n)}{{q({\bf x}_n)}}, \quad n=1,\ldots,N\nonumber
% \label{is_weights_static}
% \end{equation}
\end{itemize}
\end{enumerate}
We recall that more sampling options are also possible. In the two MIS schemes presented below, it is possible to build the UIS estimator and also to normalize the weights to create the SNIS estimator. It is important to note that, while Option 1 (s-MIS) seems a natural extension of IS to MIS, it has been shown to provide always worse performance than Option 2 (DM-MIS), quantified in the variance of the UIS estimator. In very simple examples, the difference of this variance in both cases can be of several orders of magnitude (see \cite{elvira2019generalized} for more details).
%%%%%%%%%%%%%%%%%%%%%%%%%%%%%%%%%
\subsection{Adaptive Importance Sampling and Population Monte Carlo}

Adaptive importance sampling (AIS) is an iterative procedure for the adaptation of one or several proposals. The literature of AIS is vast, specially in the last decade, and we refer to \cite{bugallo2017adaptive} for an exhaustive review. Here we focus on population Monte Carlo (PMC), a family of AIS algorithms where the adaptation is based on resampling previous weighted particles. Table~\ref{alg_pmc} describes the standard PMC algorithm~\cite{Cappe04}.
\begin{table}[!t]
	\centering
	{
	\caption{{{\color{black} Standard PMC algorithm.}}}
	
	\begin{tabular}{|p{0.95\columnwidth}|}
    \hline
		\footnotesize
		\begin{enumerate}
			\item {\bf [Initialization]}: Set $\sigma>0$, $(N,T) \in \mathbb{N}^+$. For $n=1,\ldots,N$, select the initial adaptive parameters $\bmu_n^{(1)} \in \mathbb{R}^{d_x}$ and $\bSigma = \sigma^2 \mathbf{I}_{d_x}$. %\cblue{\Real^{d_x \times d_x} }$.% = \sigma^2 \mathbf{I}$.
			\vspace*{6pt}
			\item {\bf[For $\bm t \bm= \bm 1$ to  $\bm T$]}: 
			\begin{enumerate}
				\item Draw one sample per each proposal pdf,
					\begin{equation} 
						\x_{n}^{(t)} \sim q_n^{(t)}(\x;\bmu_n^{(t)},\bSigma)
						%\label{eq_drawing_part_pmc}
					\end{equation}
					
					with $n=1,\ldots,N$.%, and $k=1,\ldots,K$. %and define the set
			%		$$
			%		\mathcal{X}_t=\left\{\x_{i,k}^{(t)}\right\}_{i=1,k=1}^{N,K}.
			%		$$
					\item Compute the importance weights,
					\begin{equation} 
						w_{n}^{(t)}=\frac{\pi(\x_{n}^{(t)})}{ q_n^{(t)}(\x_{n}^{(t)})}.
					\label{is_part_weights_pmc}
					\end{equation}
					with $n=1,\ldots,N$.
					%where $\Phi_{i,k}^{(t)}$ is a suitable function (see Section \ref{section_ppmc}).				
					\item Resample $N$ location parameters $\{{\bmu}_n^{(t+1)}\}_{n=1}^N$ from the set of $N$ weighted samples of iteration $t$.% {using the local resampling strategy}.
				%\item Adapt the proposal parameters $\{(\bmu_n^{(t+1)},\bSigma_n^{(t+1)})\}_{n=1}^N$ according to \eqref{eq:AG1}-\eqref{eq:AG2}.
\end{enumerate}
		\item {\bf [Output, $\bm t \bm =\bm T$]}: 
				Return the pairs $\{\x_{n}^{(t)}, {w}_{n}^{(t)}\}$, for
				$n=1,\ldots,N$ and $t=1,\ldots,T$.
		\end{enumerate} \\
		\hline 
\end{tabular}\label{alg_pmc}
}
\end{table}

In Step 1), the algorithm is initialized with $N$ proposals where the location parameter is set to $\bmu_n^{(1)}$ (or could be chosen randomly) and the {scale} parameter is also set to $\bSigma = \sigma^2 \mathbf{I}_{d_x}$, with $\sigma > 0$. Then, the algorithm runs for $T$ iterations as follows. In Step 2a), exactly one sample is simulated from each proposal. An importance weight is assigned to each sample in Step 2b). In Step 2c), the location parameters of next iteration are chosen by resampling the population of samples with probability proportional to the importance weight. Finally, the set of $NT$ weighted samples is returned so the classical unnormalized or self-normalized IS estimators can be built similarly to Eqs. \eqref{eq_UIS}-\eqref{eq_SNIS}.

Several PMC-based algorithms have been proposed since the publication of \cite{Cappe04}. For instance, the M-PMC in \cite{Cappe08} adapts a mixture proposal, including the weight, location, and scale parameter of each kernel of the mixture. The recent DM-PMC \cite{elvira2017improving}, adapts the location parameters of $N$ proposals while the scale parameters remain static (e.g., with Gaussian proposals, the means are adapted but not the covariance matrices). The algorithm runs also over $T$ iterations, where at each of them, $K$ samples per proposal are simulated and weighted. The $N$ adapted location parameters are resampled from the population of the $NK$ samples at time $t$.

%%%%%%%%%%%%%%%%%%%%%%%%%%%%%%%%%
\subsection{Optimization-based samplers}

The choice of a suitable proposal distribution is a key challenge in sampling algorithms and have major consequences in their performance. 
While algorithms where the proposals are parametrized by static parameters might be easier to set up, this is often suboptimal. The reason is that  the properties of the sought target are rarely known \emph{a priori}, particularly in challenging applications. Many iterative schemes for proposal adaptation have been proposed in the literature of MC samplers, with the aim of a better and faster target exploration, especially at large dimension. One of the most relevant strategies within this recent trend is the Langevin-based sampling methods. Let us assume that $\log \pi$ is continuously differentiable on $\mathbb{R}^{d_x}$. Langevin samplers are derived from discrete approximations of the continuous diffusion initially introduced in \cite{Langevin1908}. They use a gradient descent step to move the samples location in the direction of a local increase of the target. This leads to an iterative strategy called unadjusted Langevin algorithm (ULA) \cite{Roberts_MALA}: %\victor{But ULA gives you a sequence of ``samples'', right? I see what you mean, but not sure someone would see it as a ``proposal adaptation scheme where the sample is always accepted with equal weights'' (so, I would maybe present the equation first, and then connect it with IS in such way.}:
%\begin{equation}
%(\forall t \in \mathbb{N}) \quad 
%\x^{(t+1)}=\x^{(t)} + \cred{\frac{\theta^2}{2}} \nabla \log \pi(\x^{(t)}) + \cred{\theta} \boldsymbol{\omega}^{(t)},
%\label{eq:ULA}
%\end{equation}
\begin{equation}
(\forall t \in \mathbb{N}) \quad 
\x^{(t+1)}=\x^{(t)} + \frac{\epsilon^2}{2} \nabla \log \pi(\x^{(t)}) + \epsilon \boldsymbol{\omega}^{(t)},
\label{eq:ULA}
\end{equation}
where, for every $t \in \mathbb{N}$, $ \boldsymbol{\omega}^{(t)} \in \mathbb{R}^{d_x}$ is a realization of a standard Gaussian distribution and $\epsilon>0$ is the discretization stepsize. Note that the above scheme can also be interpreted as a gradient descent method perturbed with an i.i.d. stochastic error. Convergence analysis of the ULA sampler can be found for instance in \cite{Talay1990,Roberts_MALA}. As emphasized in the aforementioned works, except in very specific situations, the Markov chain generated by the ULA scheme has a unique stationary distribution which differs from the target $\pi$ (see in particular \cite{Durmus2019} for a quantification of this discrepancy). This undesirable effect is a consequence of the discretization procedure, as it is not present for the continuous Langevin diffusion \cite{Lau2007}. To overcome this limitation, the ULA can be combined with a Metropolis-Hasting (MH) strategy, based on an acceptance-reject procedure, leading to the so-called Metropolis adjusted Langevin algorithm (MALA)~\cite{roberts2002langevin}. The latter method has proved ergodic convergence, under milder assumptions on $\pi$. Moreover, its nice stability opens the door to the introduction of acceleration strategies. In particular, more sophisticated scale matrices, integrating more information (e.g., curvature) about the target~\cite{marnissi2018,Fisher_MCMC_1,Fisher_MCMC_2,qi2002hessian,XIFARA201414,roberts2002langevin,Sabanis2018}, can be adopted. Let us in particular mention the Newton MH strategy \cite{Fisher_MCMC_2,qi2002hessian}, which consists in combining an MH procedure with the stochastic Newton update:
\begin{multline}
(\forall t \in \mathbb{N}) \quad
\x^{(t+1)} = \\
\x^{(t)} + \bA(\x^{(t)}) \nabla \log \pi(\x^{(t)}) +  
 \bA^{1/2}(\mathbf{x}^{(t)}) \boldsymbol{\omega}^{(t)},
\label{eq:Newton}
\end{multline} 
where $\bA(\x^{(t)})$ is the inverse (or an approximation of it, when undefined or too complex) of the Hessian matrix $\nabla^2 \log \pi(\x^{(t)})$. {In the next section, we present our main contribution, that is a new adaptive importance sampling algorithm that integrates such Newton-based strategy within the proposal adaptation rule of an advanced PMC scheme}.

\section{Newton Population Monte Carlo }
\label{sec_slpmc}

In this section, we present the novel algorithm \emph{optimized population Monte Carlo (\acro)}, an AIS algorithm that belongs to the family of PMC algorithms (see Table \ref{alg_pmc}).
\acro~incorporates several features for an efficient adaptation of the IS proposals with the goal of approximating Eq.~\eqref{eq_integral}. The \acro~is presented in  Table \ref{PMC_framework_new}. The algorithm is initialized with $N$ proposals whose location and scale parameters are denoted with $\bmu_n^{(1)} \in \mathbb{R}^{d_x}$ and $\bSigma_n^{(1)} = \sigma^2 \mathbf{I}_{d_x}$, respectively, with $\sigma>0$, i.e., the initial scale matrices are isotropic. We denote the static parameters of the $N$ proposals as $\{\bnu_n\}_{n=1}^N$. Then, the algorithm runs for $T$ iterations, each of them divided in three steps: sampling in Step 2a), weighting in Step 2b), and adaptation in Step 2c). Finally the set of weighted samples is returned, so IS estimators can be built. In the following, we detail the steps.

\begin{table}[!t]
	\centering
	{
	\caption{{{\color{black} \acro \ algorithm.}}}
	{
	\begin{tabular}{|p{0.95\columnwidth}|}
    \hline
		\footnotesize
		\begin{enumerate}
			\item {\bf [Initialization]}: Set $\sigma>0$, $(N,K,T) \in \mathbb{N}^+$, $\{\bnu_n\}_{n=1}^N$. For $n=1,\ldots,N$, select the initial adaptive parameters ${\bm \mu}_n^{(1)} \in \mathbb{R}^{d_x}$ and $\bSigma_n^{(1)} = \sigma^2 \mathbf{I}_{d_x}$. %\cblue{\Real^{d_x \times d_x} }$.% = \sigma^2 \mathbf{I}$.
			\vspace*{6pt}
			\item {\bf[For $\bm t \bm= \bm 1$ to  $\bm T$]}: 
			\begin{enumerate}
				\item {\bf [Sampling]}: Simulate $NK$ samples as
					\begin{equation} 
						\x_{n,k}^{(t)} \sim q_n^{(t)}(\x;\bmu_n^{(t)},\bSigma_n^{(t)},\bnu_n)
						\label{eq_drawing_part_pmc}
					\end{equation}
					
					with $n=1,\ldots,N$, and $k=1,\ldots,K$. %and define the set
			%		$$
			%		\mathcal{X}_t=\left\{\x_{i,k}^{(t)}\right\}_{i=1,k=1}^{N,K}.
			%		$$
					\item {\bf [Weighting]}: Calculate the normalized IS weights as
					\begin{equation} 
						  w_{n,k}^{(t)} = \frac{\pi(\x_{n,k}^{(t)})}{\frac{1}{N}\sum_{i=1}^N  q_i^{(t)}(\x_{n,k}^{(t)})}.
					\label{is_part_weights}
					\end{equation}
					%where $\Phi_{i,k}^{(t)}$ is a suitable function (see Section \ref{section_ppmc}).
					\item {\bf [Adaptation]}: Adapt the location and scale parameters of the proposal
					\begin{enumerate}
							\item  {\bf [Resampling step]} {Resample $N$ proposals densities from the pool of $NK$ weighted samples at the iteration $t$.  The means and scales of the resampled proposals are denoted as $\widetilde {\bm \mu}_n^{(t)}$ and $\widetilde{\bSigma}_n^{(t)}$, respectively. See Section \ref{sec_scal_lang} for explicit definitions of the notations.}  %location parameters $\{\widetilde{\bmu}_n^{(t+1)}\}_{n=1}^N$ from the set of $NK$ weighted samples of iteration $t$ {using the local resampling strategy}.
				\item {\bf [Optimization step]} Adapt the proposal parameters $\{({\bm \mu}_n^{(t+1)},\bSigma_n^{(t+1)})\}_{n=1}^N$ according to \eqref{eq:AG1}-\eqref{eq:AG2}.
						\end{enumerate}	
					
				%: %$({\bm \mu}_n^{(t+1)},\bC_n^{(t+1)}) $
%$$
%\bmu_{n}^{(t+1)} = \widetilde{\bmu}_{n}^{(t+1)} + \theta_n^{(t+1)} \bA(\widetilde{\bmu}_n^{(t+1)}) \nabla \log \pi( \widetilde{\bmu}_{n}^{(t+1)} )
%$$
%and $\bC_{n}^{(t+1)} = \left(\theta_n^{(t+1)} \bA(\widetilde{\bmu}_n^{(t+1)})\right)^{1/2}$.

%%\bC_{n}^{(t+1)} & = \left(\theta_n^{(t+1)} \bA(\widetilde{\bmu}_n^{(t+1)})\right)^{1/2}.
%\end{align}
\end{enumerate}
		\item {\bf [Output, $\bm t \bm =\bm T$]}: 
				Return the pairs $\{\x_{n,k}^{(t)}, {w}_{n,k}^{(t)}\}$, for
				$n=1,\ldots,N$, $k=1,\ldots,K$ and $t=1,\ldots,T$.
		\end{enumerate} \\
		\hline 
\end{tabular}\label{PMC_framework_new}
}
}
\end{table}

\subsection{Sampling and weighting} 

 In Step 2a), at iteration $t$, each proposal is used to simulate exactly $K$ samples from it. Note that it would be possible to have a different number of samples per proposal, $K_n$, although this variation should be accordingly done with the resampling step of previous iteration $t-1$. For simplicity in the description of the algorithm, we stick to a fixed $K$.

The weighting scheme is applied in Step 2b) according to Eq. 	\eqref{is_part_weights}, and in particular, those are based on the deterministic mixture weighting scheme (DM-MIS) of Section \ref{sec_mis}. Note that these weights have been shown to provide a lower variance in UIS estimator compared to those of Eq. \eqref{is_part_weights_pmc} of the original PMC method. We can call these weights \emph{spatial} DM-MIS weights, since the proposals involved in the mixture of the denominator belong to the iteration $t$ (see other options with temporal or spatial-temporal mixtures in \cite{bugallo2017adaptive,martino2017layered}). 

\subsection{Resampling}
\label{sec_alg_resampling}

 The resampling step is the first adaptive procedure (Step 2c) in Table \ref{PMC_framework_new}) which is then followed by the optimization step. In the resampling step, we select a set of $N$ proposals, including a set of new location parameters $\{ \widetilde {\bm \mu}_n^{(t)} \}_{n=1}^N$ and the associated (inherited) scale parameters $\{ \widetilde {\bm \Sigma}_n^{(t)} \}_{n=1}^N$.  The resampling can be interpreted as a sampling procedure from one or several particle approximations of the target distribution.  

Here we present a novel resampling framework, for which existing resampling schemes are particular cases, while novel schemes can be derived (we propose one new scheme). Note that in all existing resampling schemes, only the {location parameters are resampled, while here we also resample the associated scale parameters (which is equivalent to resampling the proposals)}. In our novel framework, the set of $N$ resampled proposals with location parameters $\widetilde {\bm \mu}_n^{(t)} \triangleq {\bm \x}_{i_n^{(t)},j_n^{(t)}}^{(t)}$ and scale parameters $ \widetilde \bSigma_n^{(t)} \triangleq \bSigma_{i_n^{(t)}}^{(t)}$ are obtained by sampling (randomly) or choosing (deterministically) $N$ pairs of indexes $\{ i_n^{(t)}, j_n^{(t)}\}_{n=1}^N$. The index $i_n^{(t)}\in\{1,...,N \}$ points to the ancestor proposal which generated the sample that has been resampled, while the index $j_n^{(t)}\in\{1,...,K\}$ selects the specific sample in the set $\{\x_{i_n^{(t)},k}\}_{k=1}^K$. Note that the resampled scale parameter is selected by using only the ancestor index $i_n^{(t)}$. 
{Let us now propose three particular and interesting choices for the resampling strategies, encompassed within our versatile framework.}
%

%\begin{itemize}
	%\item 
	\paragraph*{Global resampling (GR)} %The most straightforward strategy consists in simulating $N$ times from the particle approximation $\widehat \pi^{NK}_t(\x)$, which is equivalent to the so-called \emph{global resampling} (GR) in \cite{elvira2017improving} 
	The $N$ location parameters are simulated i.i.d. from a single particle approximation $\widehat \pi^{NK}_t(\x) = \sum_{n=1}^N \sum_{k=1}^K \overline w_{n,k}\delta\left(\x - \x_{n,k}\right)$, constructed by the set of $NK$ weighted samples $\x_{n,k}$ obtained from Step~2a), and the normalized weights $\overline w_{n,k}^{(t)} = \frac{w_{n,k}^{(t)}}{\sum_{i=1}^N \sum_{j=1}^K w_{i,j}^{(t)}}$. Therefore the two indexes are simulated jointly (but each pair is independent from other pairs), leading to $n$-th pair of indexes $\{ i_n^{(t)}, j_n^{(t)} \}=\{\ell,k\}$ with associated probabilities $\overline w_{\ell,k}^{(t)}$, $\ell= 1,...,N$ and $k=1,...,K$. Note that, for such choice, the resampled particles are strongly correlated  (e.g., if one weight $\overline w_{n,k}$ dominates, all resampled particles can be identical). This scheme is closely related to the resampling step in the seminal PMC method \cite{Cappe04} and it has been recently proposed in \cite{elvira2017improving}, although in both aforementioned works it only applied to the location parameters.

 %\item 
\paragraph*{Local resampling (LR)} An alternative strategy consists in simulating exactly one sample per ancestor (i.e., proposal). In this case, alternative re-normalized weights $\widetilde w_{n,k} = \frac{w_{n,k}}{\sum_{j=1}^K w_{n,j}}$ are required, in such a way that $\sum_{k=1}^K \widetilde w_{n,k} = 1$ for all $n=1,\ldots,N$. Then, the index of the $n$-th resampled proposal, $i_n^{(t)} = n$, is chosen deterministically, while the index $j_n^{(t)}=k$ is sampled with probability $\widetilde w_{n,k}$, for each $k=1,...,K$.
 The advantage of the LR scheme is that the $N$ resampled particles are different, preserving the diversity in the exploration through $N$ paths that interact only due to the denominator in \eqref{is_part_weights}. The drawback is that it also preserves paths that are in non-relevant parts of the space. The limitations of both GR and LR strategies are closely linked to the tradeoff between particle degeneracy and path degeneracy, which is well-known in particle filtering \cite{sarkka2013bayesian,elvira2017population}.

% \item 
\paragraph*{`Glocal' resampling (GLR)} We introduce an original hybrid resampling approach, particularly tailored for the optimization-based adaptation that follows after the resampling step. %, which we call \emph{glocal resampling} (GLR): 
 The resampling step is done by following an LR step (i.e., using the $\widetilde w_{n,k}$ weights and preserving the diversity), except for the iterations with $t$ multiple of a given period parameter $\Delta \in \mathbb{N}^*$, where a GR step is performed instead. The rationale of this novel scheme is explained in Section \ref{label_resampling}.
%\end{itemize}

Finally, note that other existing schemes, such as the independent resampling (IR) of \cite{elvira2017population}, are also encompassed in this framework. 

\subsection{Optimization}
\label{sec_scal_lang}

\subsubsection{General rule}
 The second adaptive feature of our algorithm lies in Step 2c)ii). Here, in order to improve the exploration performance, we propose to adopt a Newton-based strategy for the construction of the proposal used to draw the next $KN$ samples. The proposal density for iteration $t + 1$, is modified, with a new adapted mean, given by
%\begin{equation}
%{\bm {\mu}}_{n}^{(t+1)} = \widetilde {\bm \mu}_{n}^{(t)} + \theta_n^{(t)} \bA(\widetilde{\bmu}_n^{(t)}) \nabla \log \pi( \widetilde {\bm \mu}_{n}^{(t)} ), \label{eq:AG1}
%\end{equation} 
\begin{equation}
{\bm {\mu}}_{n}^{(t+1)} = \widetilde {\bm \mu}_{n}^{(t)} + \bA(\widetilde{\bmu}_n^{(t)}) \nabla \log \pi( \widetilde {\bm \mu}_{n}^{(t)} ), \label{eq:AG1}
\end{equation} 
where $\bA(\widetilde{\bmu}_n^{(t)})$ is an SDP matrix of $\mathbb{R}^{d_x \times d_x}$. The scale matrix of the proposal is also adapted, in order to be consistent with the above location update, i.e.,
\begin{equation}
\bSigma_{n}^{(t+1)} = \bA(\widetilde{\bmu}_n^{(t)}). \label{eq:AG2}
\end{equation}
As can been seen from \eqref{eq:AG1}-\eqref{eq:AG2}, the scale matrix parameter $\bA(\cdot)$ plays an important role in our scheme, since it drives the direction and length of the adapted jump. In the following, we present our simple yet efficient strategy for the setting of this parameter. 

\subsubsection{Scaling matrix} 
Newton-based strategy amounts to integrating information of the inverse of the Hessian of $\log \pi$, in the update rule for $\bA(\widetilde{\bmu}_n^{(t)})$. However, in general cases, $\widetilde \pi$ may not be log-concave so that numerical issues can arise in the inversion of the Hessian matrix. {Furthermore, even when the inversion is well defined, one Newton iteration with unit stepsize does not necessarily yield an increase of $\log \pi$ \cite{Nocedal}. We thus propose to overcome those issues by adopting a damped Newton strategy, incorporating two specific features that aim at enforcing the numerical stability of the proposed scheme. The scaling matrix is defined as:
%\victor{read until here}
%\emilie{FIRST CASE:}
\begin{equation}
\label{eq:covadapt1a}
\bA(\widetilde{\bmu}_n^{(t)}) = \theta_n^{(t)}  \mathbf{\Gamma}(\widetilde{\bmu}_n^{(t)}),
\end{equation}
with
\begin{equation}
\label{eq:covadapt1b}
\mathbf{\Gamma}(\widetilde{\bmu}_n^{(t)}) = 
\begin{cases}
\left(- \nabla^2 \log \pi(\widetilde{\bmu}_n^{(t)})\right)^{-1}, & \text{if} \; \nabla^2 \log \pi(\widetilde{\bmu}_n^{(t)})\succ 0,\\
{\widetilde \bSigma_{n}^{(t)}}, & \text{otherwise}.
\end{cases}
\end{equation}
Otherwise stated, the covariance of the $n$-th proposal is set by using second order information when it yields to a definite positive matrix; otherwise, it inherits the covariance of the {$i_n^{(t)}$}-th proposal that generated the sample. 
%\emilie{SECOND CASE:}
% \begin{equation} 
% \label{eq:covadapt2}
% \bA(\widetilde{\bmu}_n^{(t)}) =
% \begin{cases}
% \theta_n^{(t)} \left(- \nabla^2 \log \pi(\widetilde{\bmu}_n^{(t)})\right)^{-1} & \text{if} \; \nabla^2 \log \pi(\widetilde{\bmu}_n^{(t)})\succ 0\\
% \sigma^2 \mathbf{I}_{d_x} & \text{otherwise}.
% \end{cases}
% \end{equation}
 Moreover, we introduced $\theta_n^{(t)}\in (0,1]$, which is a stepsize tuned according to a backtracking scheme in order to avoid the degeneracy of the Newton iteration, and thus of our adaptation scheme. Starting with unit stepsize value, we reduce it by factor $\tau =1/2$ until the condition below is met: 
\begin{equation}
\pi\left(\widetilde{\bmu}_n^{(t)} + \bA(\widetilde{\bmu}_n^{(t)}) \nabla \log \pi( \widetilde {\bm \mu}_{n}^{(t)} )\right) \geq \pi\left(\widetilde{\bmu}_n^{(t)}\right) \label{eq:backtrack}.
\end{equation}
}

\subsection{Related works in the literature}
 
The PMC algorithms perform the adaptation of the proposals via a resampling scheme. 
%
% There are few exceptions, being \cite{Cappe08} the most relevant one (there is not a resampling step but a mixture adaptation)
%
This step can be viewed as a stochastic procedure, since it is based on a multinomial resampling with replacement. However, since the set of proposals in the next iteration is parametrized by the resampled particles, this procedure can be alternatively seen as an implicit optimization procedure (in general, this perspective is not mentioned in the literature). In this paper, we propose, for the first time up to our knowledge, an explicit optimization procedure incorporated within the adaptation part of the algorithm, more precisely after the resampling step is done. Moreover, we design a suitable resampling step that allows the optimization step to exploit the benefits of stochastic and deterministic adaptation.

%\cgreen{Defend the novelty of our PMC scheme (weighting/resampling part)}

The introduction of optimization-based rules for improving the exploration properties of other AIS methods, not belonging to the PMC family, has been explored in the recent works~\cite{elvira2015gradient,schuster2015gradient,fasiolo2018langevin,akyildiz2021convergence}. In \cite{schuster2015gradient}, the authors propose a gradient descent with decreasing stepsize update for the location parameters, while the covariance update relies on the calculation of the empirical covariance of the past samples. Moreover, there is only one proposal. % and not a sense of continuity of the proposal trajectories.} %\emilie{I do not understand `a sense of continuity'}%(it can be seen as the AMIS algorithm \cite{CORNUET12} where the sample is simulated via resampling instead of the traditional momement-matching of AMIS, and then a gradient step is peformed)}. 
In GAPIS \cite{elvira2015gradient}, the location parameters are updated according to a Newton step, \cblue{with the stepsize remaining static, while the covariance is adapted using the Hessian of $-\log \pi$.} Finally, in NIMIS and LIMIS \cite{fasiolo2018langevin}, a temporal mixture is constructed, in the spirit of AMIS \cite{CORNUET12} but using a mixture that increases the number of components with the iterations (instead of remembering the whole mixture simply for the calculation of the importance weight). In LIMIS, the location parameters move along a gradient direction while the covariance adaptation relies on second-order approximation of the target, both updates being evaluated using Runge-Kutta numerical integration. Up to our knowledge, the use of a Newton-based adaptation for both location/covariance parameters, has never been considered in PMC literature.

%Several strategies have been investigated for the construction of the scaling matrix in the case of ULA and its Metropolis version MALA, relying on Hessian~\cite{Newton_MCMC_1,Newton_MCMC_2,Qi2002}, Fisher metric \cite{Fisher_MCMC_2} or majorization-minimization strategy \cite{marnissi2018}. The aforementioned works share our idea to (i) incorporate second-order information within the update scheme so as to better map the local curvature of the target, (ii) use a scalar stepsize to control the acceptance rate/convergence stability of the chain. Although initially introduced for MCMC samplers, it is worth noting that the above Langevin and Hessian based procedures for proposal adaptation have also been used recently in the context of PMC algorithms \cite{schuster2015gradient,fasiolo2018langevin,elvira2015gradient}. As mentioned in our introduction, in such case, the proposal adaptation of location parameters relies on a resampling strategy, possibly combined by an optimization (e.g., gradient descent \cite{schuster2015gradient}, Newton \cite{elvira2015gradient}) step, while the covariance proposal might be either static \cite{elvira2015gradient} or adapted \cite{fasiolo2018langevin}.  } \emilie{I copy this here, need to be rewritten !}.

\section{Discussion}
In the following, we discuss the key elements of the novel \acro\ algorithm. 

\subsection{Importance weights}
In \acro \, $K$ samples are simulated from $N$ proposals at each iteration. Then, the importance weights are computed in Eq. \eqref{is_part_weights}. First, note that these weights do not follow the standard functional consisting on the ratio between the target and the proposal distributions, e.g., the sample $\x_{n,k}^{(t)}$ is simulated from the $n$-th proposal but in the denominator of $w_{n,k}^{(t)}$, the whole mixture $\sum_{i=1}^N  q_i^{(t)}(\x)$ is evaluated (instead of just $q_n^{(t)}$). This alternative choice for the weights, called balance heuristic \cite{Veach95} or deterministic weights \cite{Owen00}, has been shown to be unbiased and even more, to provide IS estimators with reduced variance \cite{elvira2019generalized}. The benefits of such alternative weights go beyond the superior performance of the estimators, and provide advantages in the resampling adaptation, compared to the standard weights. The reason is that, when evaluating the whole mixture in the denominator, the importance weight captures the mismatch between the target and the whole mixture of proposals at the iteration $t$ (and not only the particular proposal that generated the sample). The resampling stage done with these weights allows to over-sample regions that are under-represented (see next section).

Finally, note that a {mixture} with the whole set of proposals $\{q_n^{(\tau)} \}_{1 \leq n\leq N,1 \leq \tau \leq t}$, could be placed in the denominator of the importance weight, in the spirit of the AMIS algorithm \cite{CORNUET12}. We have however discarded this option as it would increase the computational complexity, particularly when $T$ is large \cite{bugallo2017adaptive}.
%\cgreen{Unbiasedness and variance reduction}
%\victor{Not really new. Not sure how to improve/extend it for this paper.}
%\victor{We have considered full DM that can reduce the variance (like in AMIS) but we decide to not implement it due to extra computational complexity.}

\subsection{Resampling schemes}
\label{label_resampling}
PMC algorithms adapt the set of proposals via a resampling step. In the seminal PMC algorithm from \cite{Cappe04}, the resampling is done at each time step using the standard weights. New resampling schemes have been proposed in \cite{elvira2017improving,elvira2017population}. Note that by \emph{}resampling scheme}, we do not only refer to the way the sampling of the indexes that will be replicated is done (as in \cite{Douc05,li2015resampling}). In PMC algorithms, the samples are not i.i.d., and hence it is possible to enforce different adaptation behaviors. It is important to note that unlike in adaptive MCMC methods, where modifying the adaptation can endanger the convergence of the method, in AIS the needed assumptions are milder, since the validity of the estimators is ensured by the importance weights. 

In \acro \, we propose two possible resampling schemes that are designed so as to exploit the \emph{optimization step} (see the step 2)c)ii) in Table \ref{PMC_framework_new}). The local resampling (LR), proposed in \cite{elvira2017improving} is particularly suitable for the optimization step that follows the resampling. As described in Section \ref{sec_alg_resampling}, the LR scheme ensures that one (and only one) sample simulated from the $n$-th proposal survives. This is advantageous for the Newton-based step that follows the resampling, since in practice there are $N$ \emph{chains} or threads, corresponding to the resampled particle among the set $\{\x_{n,k}^{(t)}\}_{k=1}^K$ for each $n=1,...,N$, that is later adapted using the geometry of the target. It is interesting to see that these $N$ \emph{chains} interact only in the importance weights calculation (because the whole mixture is in the denominator, as explained in the previous section). \cblue{This interaction is very effective in combination with the other features of \acro, since it can give a lower weight to specific samples that are in over-populated areas (i.e., other proposals are covering that region), even if this part of the target has high probability mass. Therefore, even if the location parameter of each proposal is independently displaced to a region of higher probability mass via the Newton step with the risk of overpopulating the region around the mode(s), the resampling step re-balances the proposals at each iteration to approximate the target as a mixture.} In \acro, we have discarded the use of the standard global resampling (GR) \cite{Cappe04,elvira2017improving} since it is known to reduce the diversity, endangering the exploration of the target (after the GR step, all resampled particles can come from the same proposal, and even more, can be exactly the same).

Moreover, in \acro \ we present another variant called \emph{glocal} resampling (GLR). The GLR scheme can be seen as a modified LR scheme, where, at every $\Delta \in \mathbb{N}^*$ iterations, a GR step is performed instead of an LR one. The rationale is to preserve for most iterations the diversity in the exploration of each proposal (with a mild interaction in the denominator of the importance weights). Periodically, every $\Delta$ iterations, the GR step enforces a stronger cooperation among paths, killing those who are in irrelevant parts of the space, and replicating those who are more promising (which allows for a more exhaustive local exploration in the next iterations). The GLR strategy keeps clear ties with the adaptive resampling \cite{del2012adaptive}, allowing to find a good balance between an accurate local approximation of the target and a global exploration.
%\victor{GLOCAL, if LOCAL does not converge}
%\victor{LOCAL would converge to the mode but does not balance the global weight (in the mixture) of each proposal/mode. Local with the mixture in the denominator cooperates a bit, but fails to identify the relative importance of each mode.}

\subsection{Newton-based adaptation}

\subsubsection{Improvement w.r.t. Newton scheme}
In the optimization step, a straightforward approach may be simply choosing the scaling metric by relying on the information of the Hessian of $\log \pi$. In this approach, we might set $\bA(\widetilde{\bmu}_n^{(t)})$ as the invert of $\nabla^2 \log \pi(\widetilde{\bmu}_n^{(t)})$. In such a way, \eqref{eq:AG1} would read as one Newton iteration applied to the maximization of function $\log \pi$, and initialized at $\widetilde{\bmu}_n^{(t)}$. However, there are two drawbacks for the Newton optimization method, namely (i) the requirement for convexity of $- \log \pi$ for proper definition of the iteration, (ii) the local convergence behavior, i.e., convergence only when initialized ``sufficiently close'' to a mode. We thus propose two controlling rules, to avoid these difficulties. First, our proposed scheme in \eqref{eq:covadapt1b} introduces a test, taking into account the fact that $\log \pi$ might not be necessarily log-concave at $\widetilde{\bmu}_n^{(t)}$, so that $\nabla^2 \log \pi(\widetilde{\bmu}_n^{(t)})$ might be non invertible. {We take here advantage of the trajectory tracking inherent to the AIS method, by re-using the past scaling matrix from the particle ancestors.} {This is particularly advantageous, since the samples are in general close from the location parameter of the density where they were simulated. \cblue{An alternative to this approach would be inheriting the scale parameter of the proposal (at iteration $t$) that is closer to the sample, as it is done for instance in \cite{fasiolo2018langevin}. This may increase the performance without increasing the complexity if the proposals are Gaussian pdfs and the criterion is based on the Mahalanobis distance (since these distances are implicitly computed in the denominator of the importance weights). Another more ambitious scheme would be doing the same, but considering the set of whole set of $Nt$ present and past proposals. While these alternative may capture better the second order information of the target in the neighborhood of the sample, in general they would translate into an increase of complexity. This is highly related to the well-known increase of complexity in AMIS algorithm when the number of iterations grow, with efficient versions of the algorithm trying to alleviate this issue, e.g., the EAMIS \cite{el2019efficient}.} Hence, \acro \ either accounts for the second-order information at the sample location or inherits a more stable one from a close location.} %\emilie{I want to emphasize how convenient and elegant it is, to use inheritance of past scale, can you check and maybe reformulate}. 
Second, a stepsize is introduced in \eqref{eq:covadapt1a}, which is computed following a simple backtracking procedure. The idea is to constrain the Newton step within a region in which we believe that the second order model, inherent from the Newton approximation on $\log \pi$ is reliable, using iterative trials for the stepsize. If a notable increase of $\log \pi$ is gained, then the model is believed to be a good representation of the original objective function. If there is not significant improvement, the model is considered invalid, and a new step is tried. It is worth noting that the fulfillment of the descent condition \eqref{eq:backtrack} in finite time is ensured under mild assumptions on $\log \pi$ (e.g., Lipschitz differentiability, see \cite{Nocedal}). Moreover, under the same assumptions, the unit stepsize satisfies \eqref{eq:backtrack} as soon as $\widetilde{\bmu}_n^{(t)}$ is sufficiently close to a local maximum of $\log \pi$ \cite{Nocedal}. Otherwise stated, the classical (fast) Newton move of the location parameters is retrieved as soon as the particles get close to the modes of the target. {More sophisticated approximations for the Hessian (or its inverse) may be desirable when exploring very large scale multimodal distributions. {Low-rank \cite{Newton_MCMC_1}  or majorizing \cite{marnissi2018} approximations, proposed in the context of Langevin-based MCMC, appear as appealing alternatives. However, it is not straightforward to incorporate those approaches in the proposed \acro \ scheme, while keeping the versatility of the algorithm. The exploration in high-dimensional problems may be improved by imposing an isotropic/diagonal structure in the covariance matrix  \cite{fernandez2015probabilistic}, and fitting it through a particle approximation  via importance sampling, at the expense of reducing the efficiency of the estimators in highly correlated target distributions.}%\emilie{not sure i understand your reformulation. I would say they could NOT, actually ?} 
%\victor{I like paragraph, with the two mechanisms. I wonder if this should be more highlighted in abstract, intro, and Sec. III. Even if ``obvious'' for you, these well-grounded state-of-the-art mechanisms from optimization community can be sold as novelty and serious research in the AIS context.}

%Let us remark that, despite the complexity increase when compared to a standard first-order scheme that would be obtained by setting $\bA(\cdot)$ proportional to identity matrix, such Newton-based strategy for proposal adaptation has been observed to yield very satisfactory performance in the context of MCMC \cite{Newton_MCMC_1,Newton_MCMC_2}. \cred{It has also been used for the covariance adaptation in PMC methods \cite{fasiolo2018langevin,elvira2015gradient}, again with great performance.} \emilie{maybe not interesting comment and even redundant with previous sections ?} \victor{I am not sure if I understand the paragraph. If you do not see that it adds value, maybe we can remove it.}

%Here, we propose to simply choose the scaling metric in our Langevin-based scheme relying on the information of the Hessian of $\log \pi$:
%\begin{equation}
%\bA(\widetilde{\bmu}_n^{(t)}) = \theta_n^{(t)}  \left(- \nabla^2 \log \pi(\widetilde{\bmu}_n^{(t)})\right)^{-1}, \label{eq:newton}
%\end{equation}
%with $\theta_n^{(t)} \in (0,1)$ is a controlling stepsize and $\nabla^2 \log \pi$ the Hessian of $\log \pi$, . 

\subsubsection{Connection with scaled Langevin dynamics}
{
%\victor{The next part: instead of saying that it is different from Langevin, would it better help to say that there are interesting connections (that help to understand our good performance) but with some differences? Just some idea, maybe it is not as similar as we discussed.}
%{Finally, let us point out that 
Our scaled gradient adaptation scheme \eqref{eq:AG1}-\eqref{eq:AG2} keeps interesting connections with the discretized version of the scaled Langevin diffusion, mentioned for instance in \cite{roberts2002langevin,marnissi2018} in the context of MCMC sampling. {This discretized Langevin diffusion can be expressed as (using similar notations as in \eqref{eq:ULA}):}
\begin{equation}
(\forall t \in \mathbb{N}) \quad 
\x^{(t+1)}=\x^{(t)} + \epsilon^2 \bb(\x^{(t)}) + \epsilon \bA^{1/2}(\mathbf{x}^{(t)}) \boldsymbol{\omega}^{(t)}.
\label{equLangevinScaled}
\end{equation}
Hereabove, $\bb: \mathbb{R}^{d_x} \to \mathbb{R}^{d_x}$ is the so-called drift term such that, for every $1 \leq i \leq d_x$ and every $\x \in \mathbb{R}^{d_x}$,
\begin{multline}
b_i(\x)=\frac{1}{2} \sum_{j=1}^{d_x} A_{i,j}(\x) \dfrac{\partial \log \pi(\x)}{\partial x_j} + \\ \vert \bA(\x) \vert^{\frac{1}{2}}  \sum_{j=1}^{Q} \dfrac{\partial}{\partial x_j}\left(  A_{i,j}(\x) \vert \bA (\x) \vert^{-\frac{1}{2}} \right), 
\label{equation_diff}
\end{multline}
with $\bA(\x) \in \mathbb{R}^{d_x \times d_x}$ a symmetric definite positive (SDP) matrix with determinant $\vert \bA(\x) \vert$. A typical approximation, adopted in \cite{marnissi2018}, consists in ignoring the second term in \eqref{equation_diff}, leading to the simplified sampling scheme:
\begin{multline}
(\forall t \in \mathbb{N}) \quad
\x^{(t+1)}= \\
\x^{(t)} + \cblue{\frac{\epsilon^2}{2}} \bA(\x^{(t)}) \nabla \log \pi(\x^{(t)}) +  
%(2 \theta^{(t)})^{1/2} 
\cblue{\epsilon}
\bA^{1/2}(\mathbf{x}^{(t)}) \boldsymbol{\omega}^{(t)},
\label{eq:3mh} 
\end{multline}
where \cblue{$\epsilon$ is a positive stepsize and $\{ \bA(\x^{(t)})\}_{t \geq 0}$ are positive symmetric definite positive scale parameters, possibly varying over the discrete time iterates indexed by $t \in \mathbb{N}$}. Ergodicity of the chain generated by \eqref{eq:3mh}, combined with a Metropolis-Hasting step, was established in~\cite{marnissi2018}, for a large class of choices for \cblue{$\{\bA(\x^{(t)}) \}_{t \geq 0}$ and $\epsilon$}. It is noticeable that our optimization-based adaptation scheme in Eq. \eqref{eq:AG1}-\eqref{eq:AG2} is closely related to \eqref{eq:3mh}, \cblue{identifying $\theta^{(t)}$ with $\frac{\epsilon^2}{2}$}. \cblue{Note that no factor $2$ is present in our covariance adaptation rule in Eq.~\eqref{eq:AG2}}, in a similar fashion as in the Newton MH sampler from \cite{Fisher_MCMC_2,qi2002hessian}. In this way, the scale parameter of the proposal adapts, in a robust way, to the curvature of the target distribution, {as we show in the next toy example.} %\victor{I added this last sentence to explain intuitively why, and to connect with the next subsection. Feel free to modify or extend.} %The AIS algorithms actually aim at reconstructing the target distribution with the mixture of $N$ proposals (see \cite{bugallo2017adaptive} for a discussion), and hence our choice of covariance adaptation \eqref{eq:AG1}-\eqref{eq:AG2} aims at locally recreating the curvature of the target.
}

 \subsection{Toy example}
{
We illustrate the behavior of \acro \ {along iterations on a simple example where the target is a mixture of bivariate Gaussian distributions,} with means $[-5,-5]^\top$ and $[6,4]^\top$, covariances $[0.25,0 ; 0,0.25]$ and $[0.52,0.48;0.48,0.52]$, and mixture weights $0.7$ and $0.3$, respectively. 
We run $T = 10$ \acro \ iterations with $(N,K) = (50,20)$, and resampling schemes LR and GLR (with period $\Delta = 5$). We initialize {the location parameters of the proposals} randomly in the square $[-5,5] \times [-5,5]$, and the initial covariance is set to $\bSigma_n^{(1)} = \bI_2$. \cblue{We also run the GR-PMC and LR-PMC algorithms \cite{elvira2017improving}, the AMIS algorithm \cite{CORNUET12}, and the GAPIS algorithm\cite{elvira2015gradient} algorithms, all with the same parameters for a fair comparison.} We display in Fig.~\ref{fig:toy_example_good_sigma} the evolution {along the iterations of the proposals, including the location parameters (black dots) and scale parameters (green ellipses). We also show the two marginal pdfs of the target distribution (blue line) and the equally weighted mixture of proposals (red line).}
We notice that \acro \ algorithm moves rapidly the proposal locations to the two modes. Moreover, it fits the  scale parameters of the proposals to the scale parameters of each mode (depending on the part of the space where the proposal is located).} The target is thus very accurately estimated, in {few iterations}, as can be seen in the 2D plots as well as in the marginals. In contrast, both GR-PMC and LR-PMC schemes struggle to reach {a reasonably good target approximation.} We also observe the benefits offered by {GLR, our novel resampling scheme, as it can be easily noticed the improved performance of \acro \ w.r.t. the LR variant.} This is particularly visible between $t=5$ and $t=6$, as $t=6$ corresponds to the first (periodic) callback to the GR resampling {(see Sec. \ref{sec_alg_resampling} for more details). \cblue{GAPIS discovers both modes, but since the step-size is not adapted in this algorithm, after $t=10$ iterations the proposals are still in between the initialization and the asymptotic value (that would be close to the modes of the targets). Interestingly, GAPIS incorporates a repulsion behavior between proposals that has certain parallelism with the resampling step in O-PMC (due to the DM-MIS weights). The advantage of O-PMC is that this implicit repulsion does not require extra parameters. In this example, AMIS fails to discover one of the modes due to an initialization that was designed to be challenging for all algorithms. A different initialization (or a much higher initial variance of the proposals) may help the algorithm at the expense of being more inefficient and taking more iterations to converge.}

%\emilie{+ add comments on GAPIS/AMIS results}. \victor{add discussion with parallelism with  GAPIS and why not repulsion is needed.}

%\emilie{I omitted GAPIS : the results were embarrassing, as it was obvious that a simple stepsize tuning would make it much better in location adaptation.}

%\victor{I like a lot the experiment, although not many insights maybe for all those plots? I think that in general, the algorithms that did not learn much in iteration 2, they don't learn much (so t=5 and 20 may be redundant (except in GAPIS)). For the moment I do not see how to do this better. I would select only one sigma maybe.}

\begin{figure*}[htp]  
\centering 
\begin{tabular}{@{}c@{}c@{}c@{}c@{}}
\includegraphics[width=0.5\columnwidth]{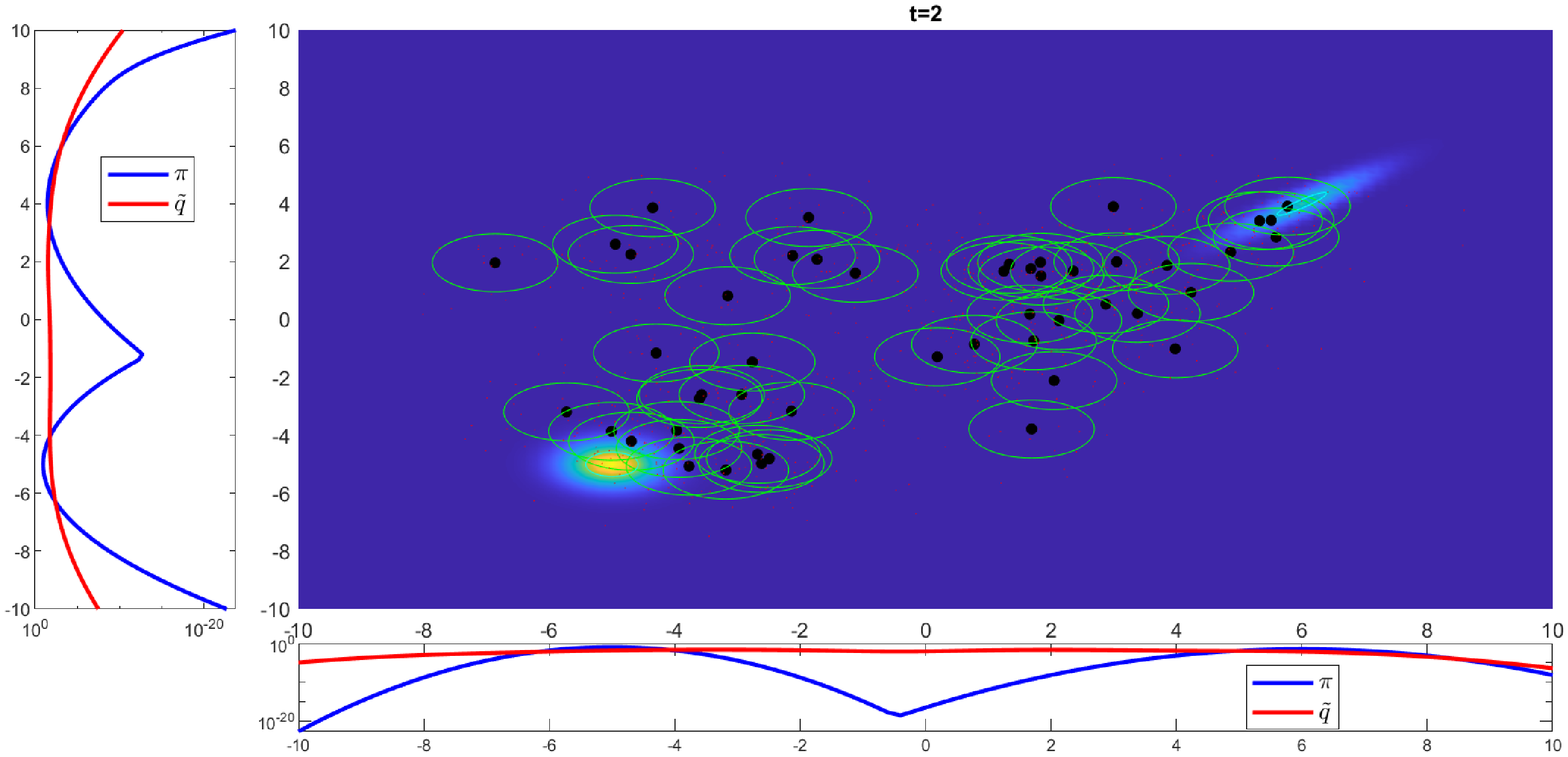} & \includegraphics[width=0.5\columnwidth]{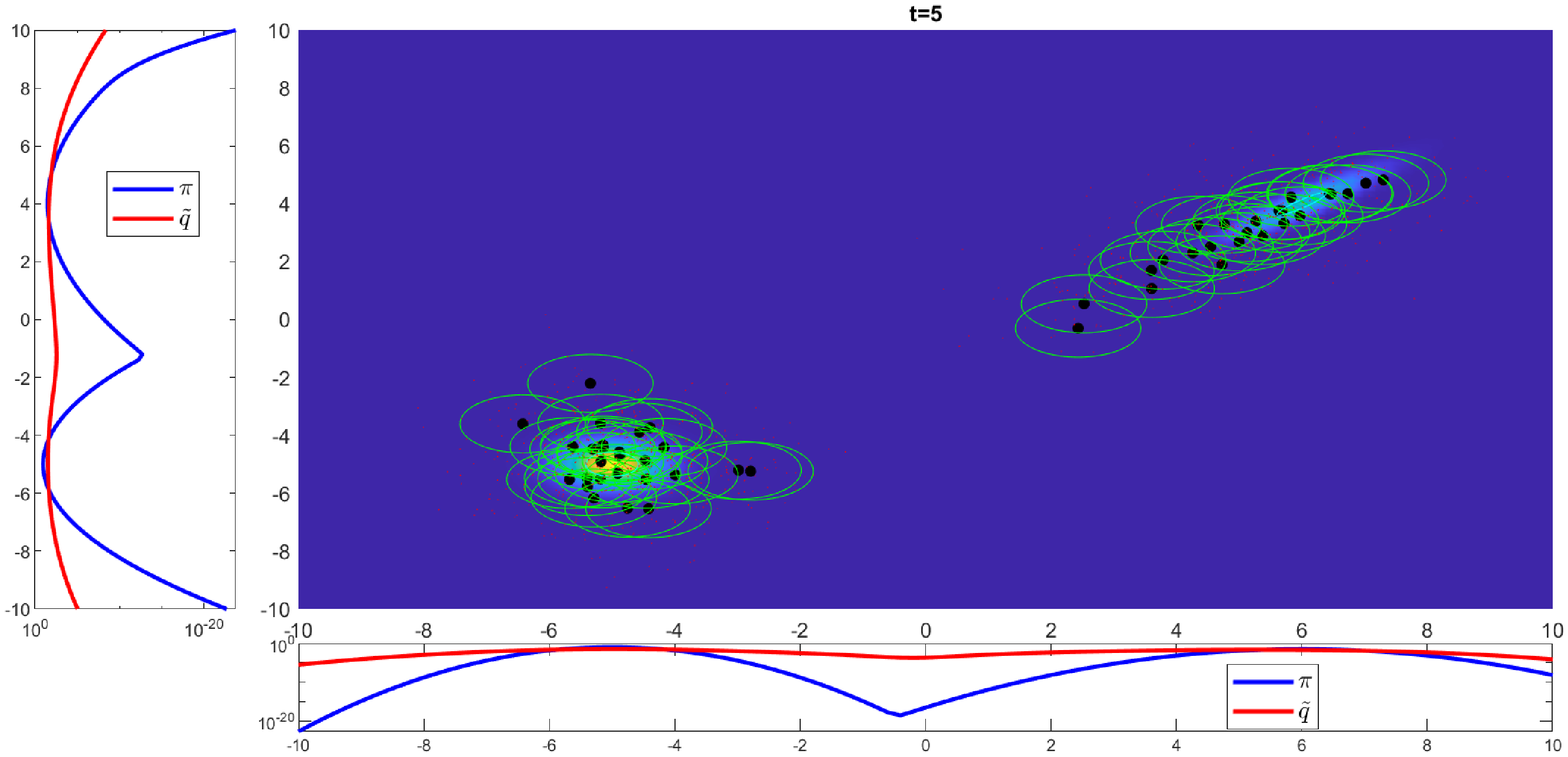} & \includegraphics[width=0.5\columnwidth]{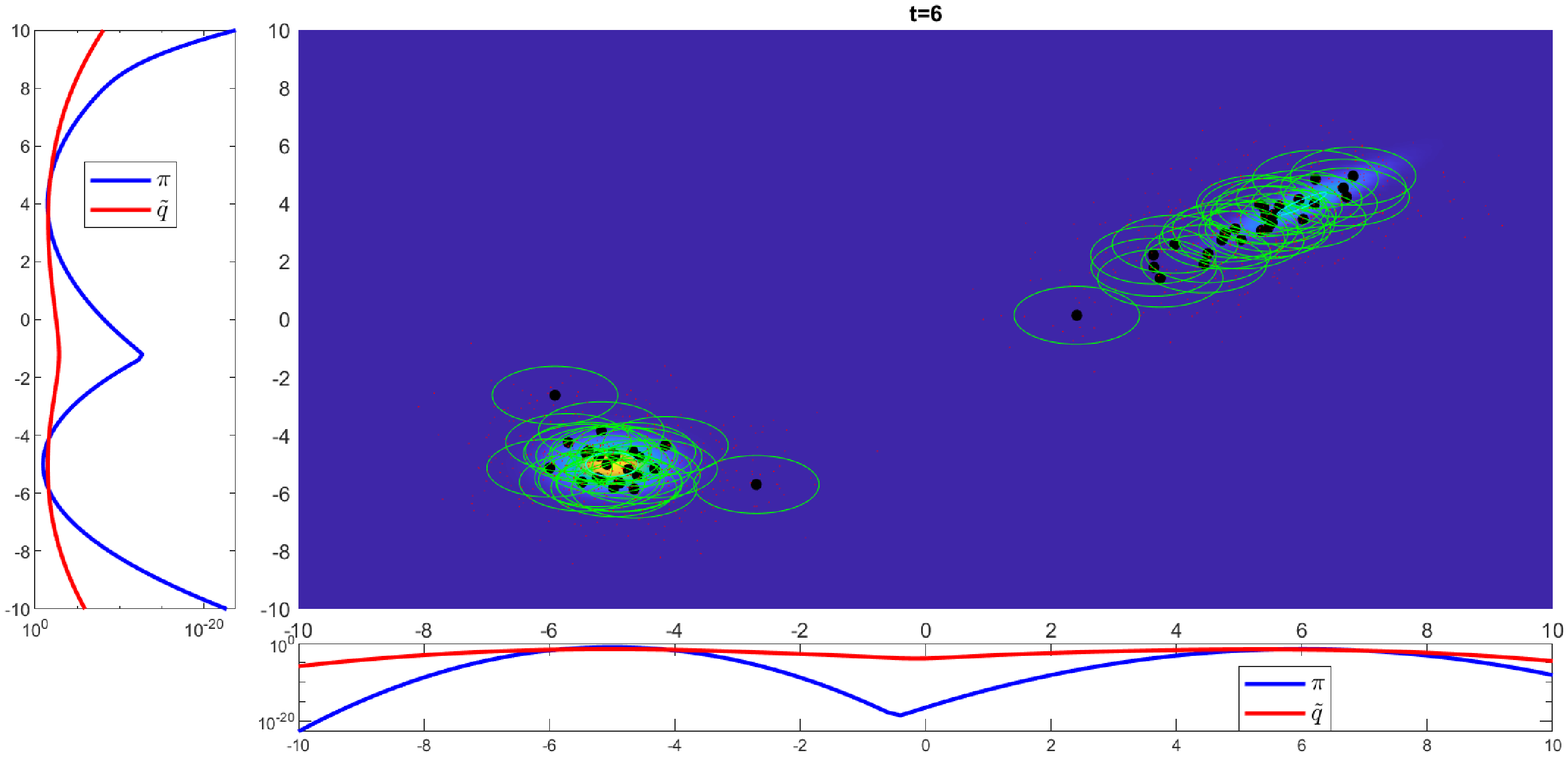} & \includegraphics[width=0.5\columnwidth]{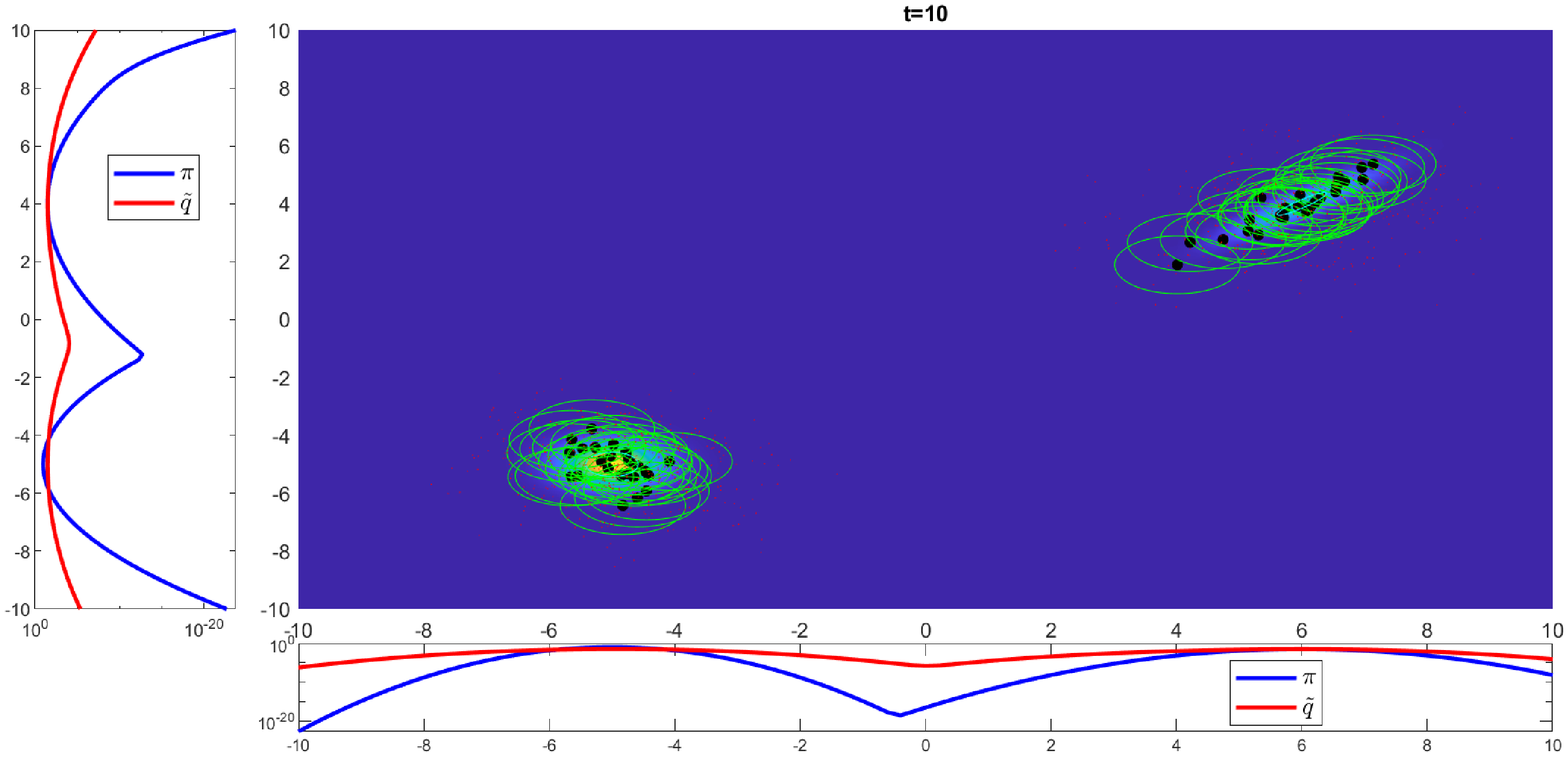}\\
LR-PMC ($t=2$) & LR-PMC ($t=5$) & LR-PMC ($t=6$) & LR-PMC ($t=10$)\\
\includegraphics[width=0.5\columnwidth]{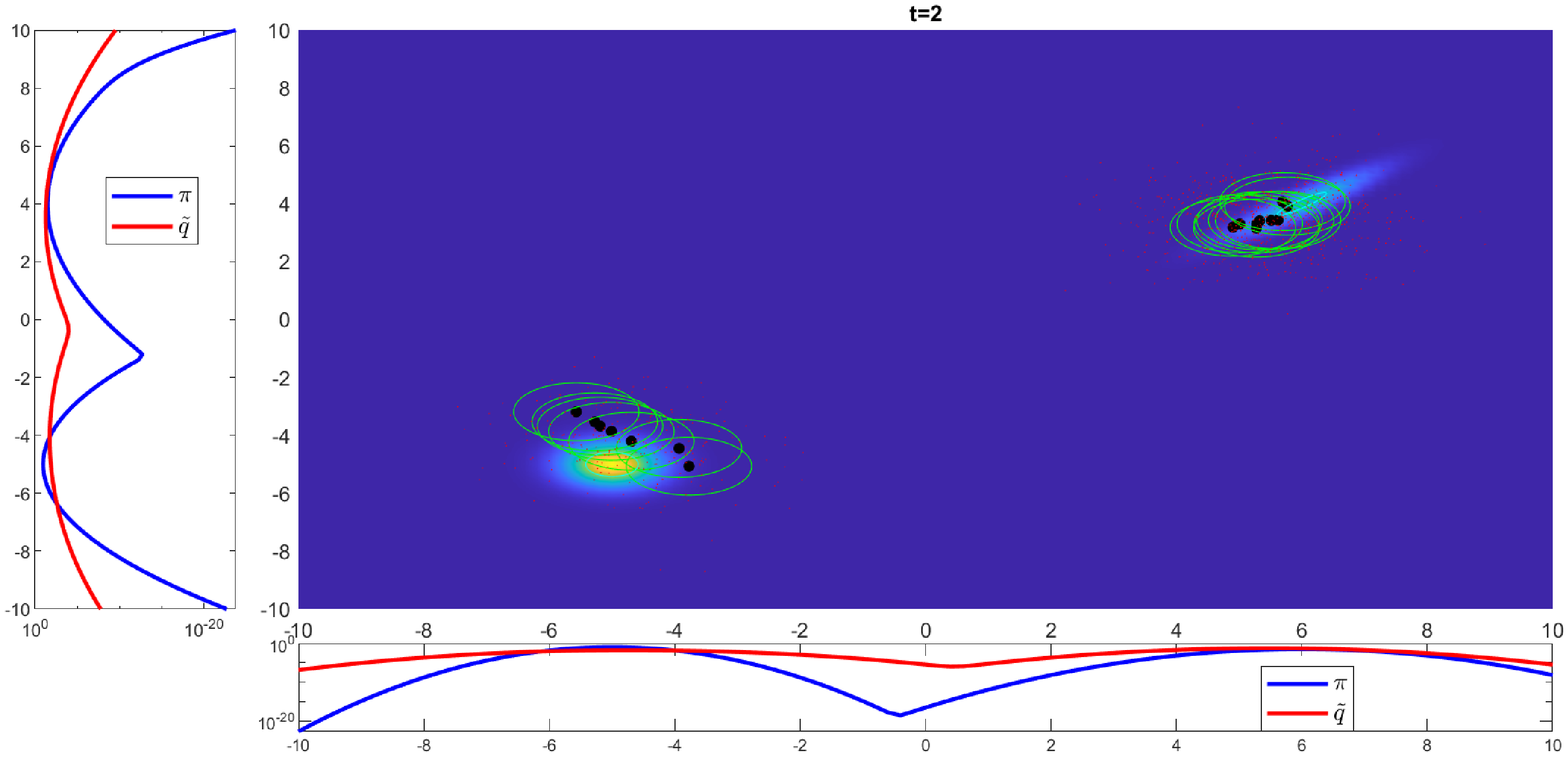} & \includegraphics[width=0.5\columnwidth]{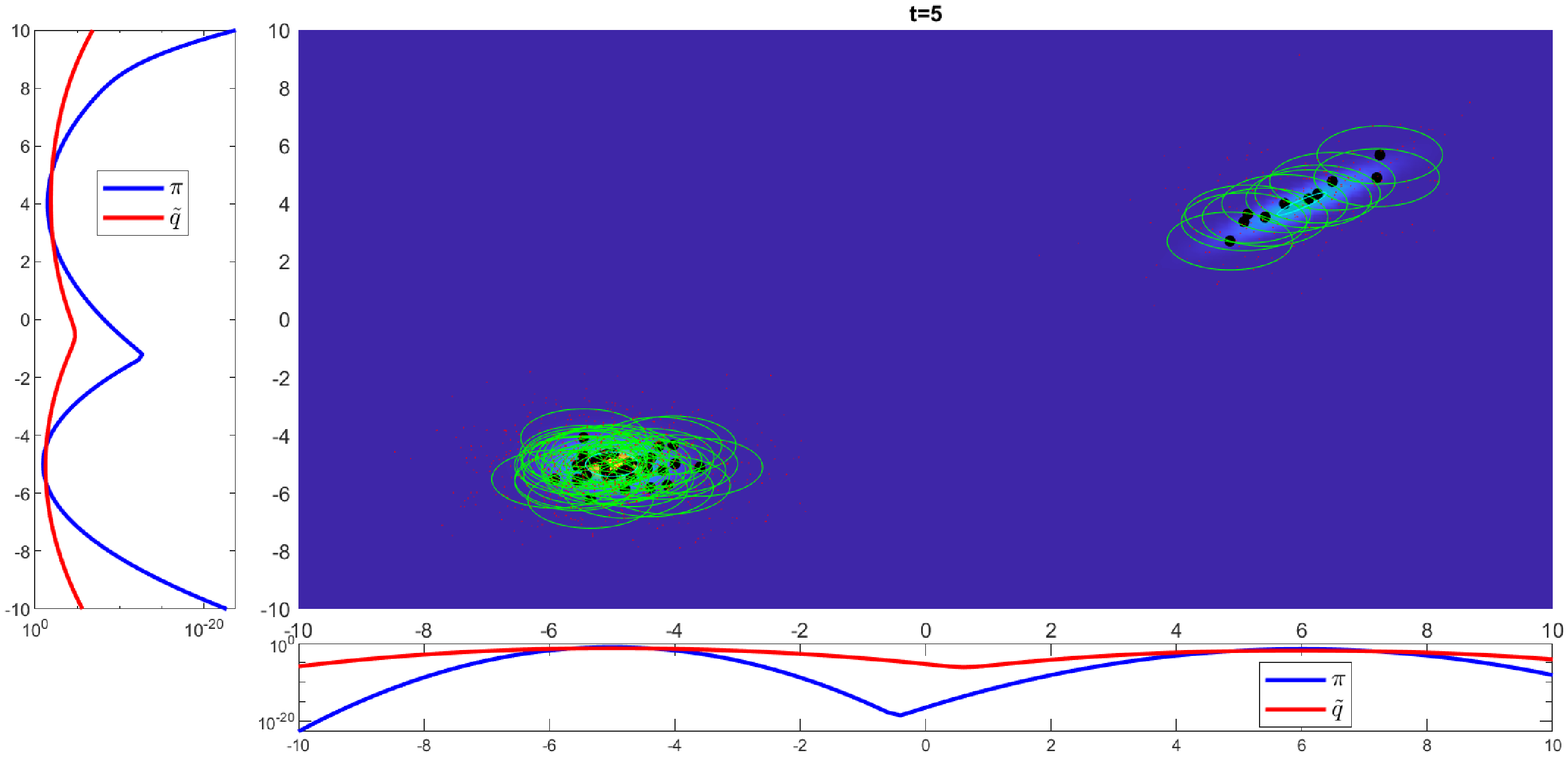} & \includegraphics[width=0.5\columnwidth]{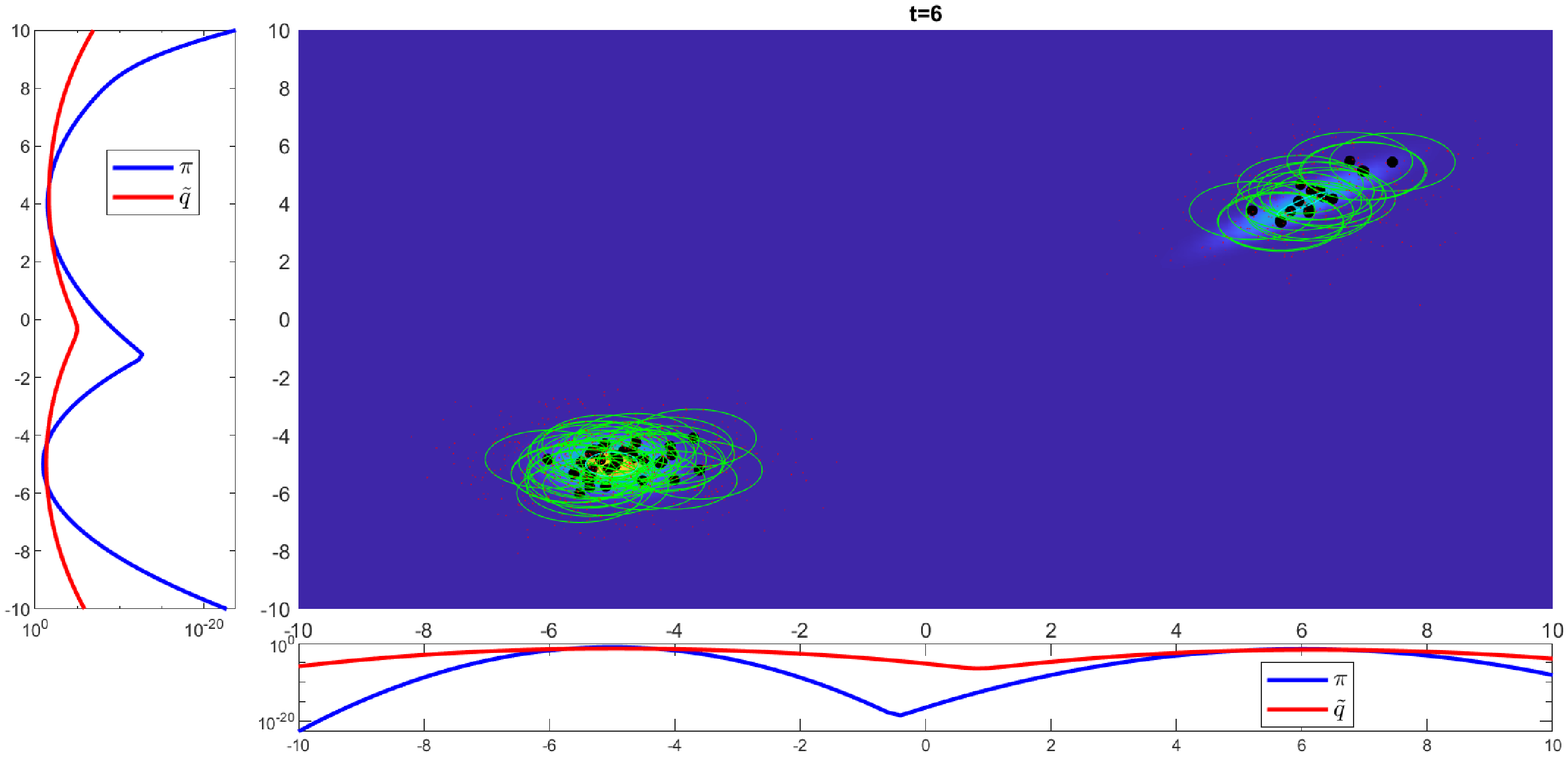} & \includegraphics[width=0.5\columnwidth]{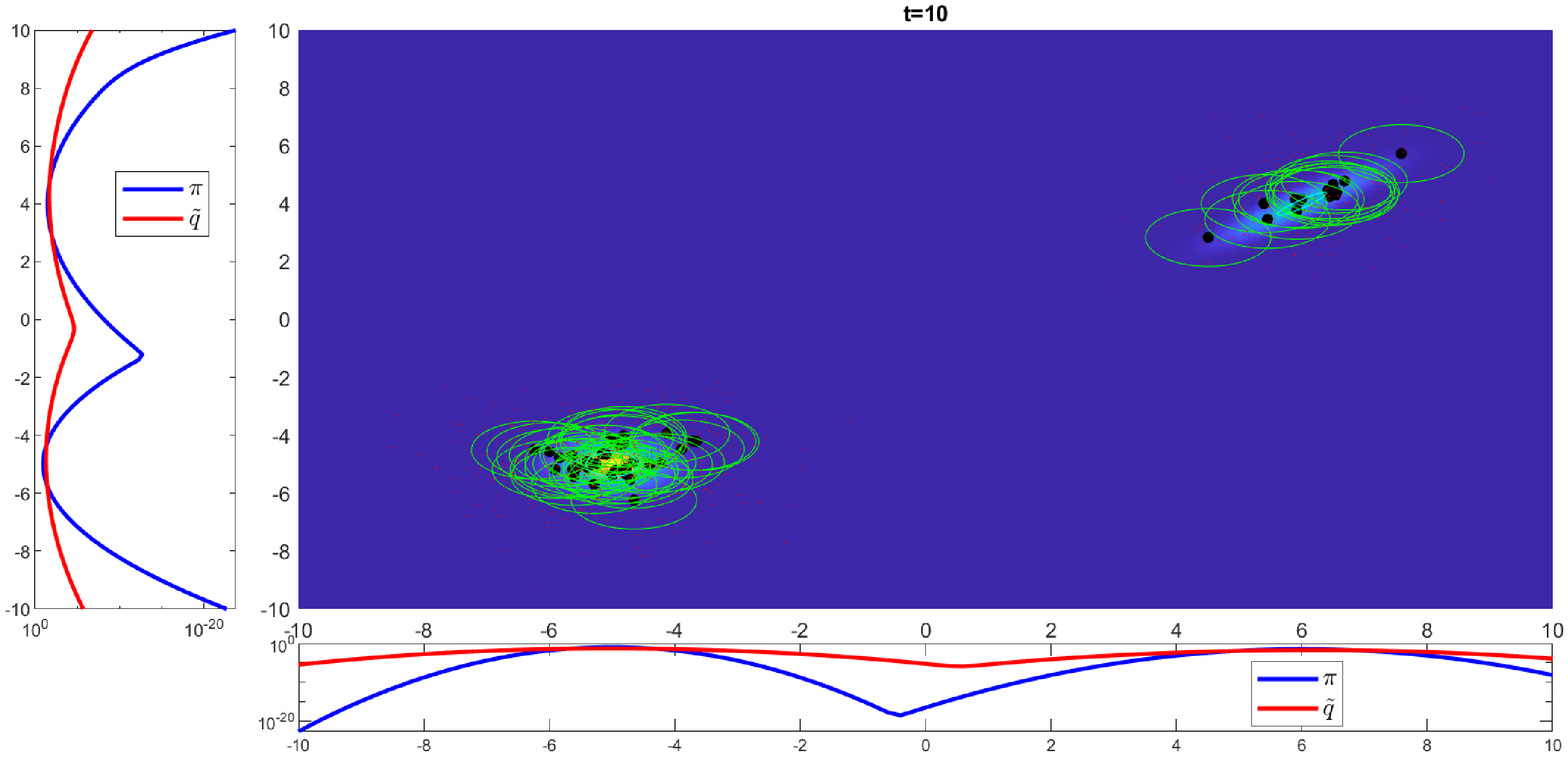}\\
GR-PMC ($t=2$) & GR-PMC ($t=5$) & GR-PMC ($t=6$) & GR-PMC ($t=10$)\\
\includegraphics[width=0.5\columnwidth]{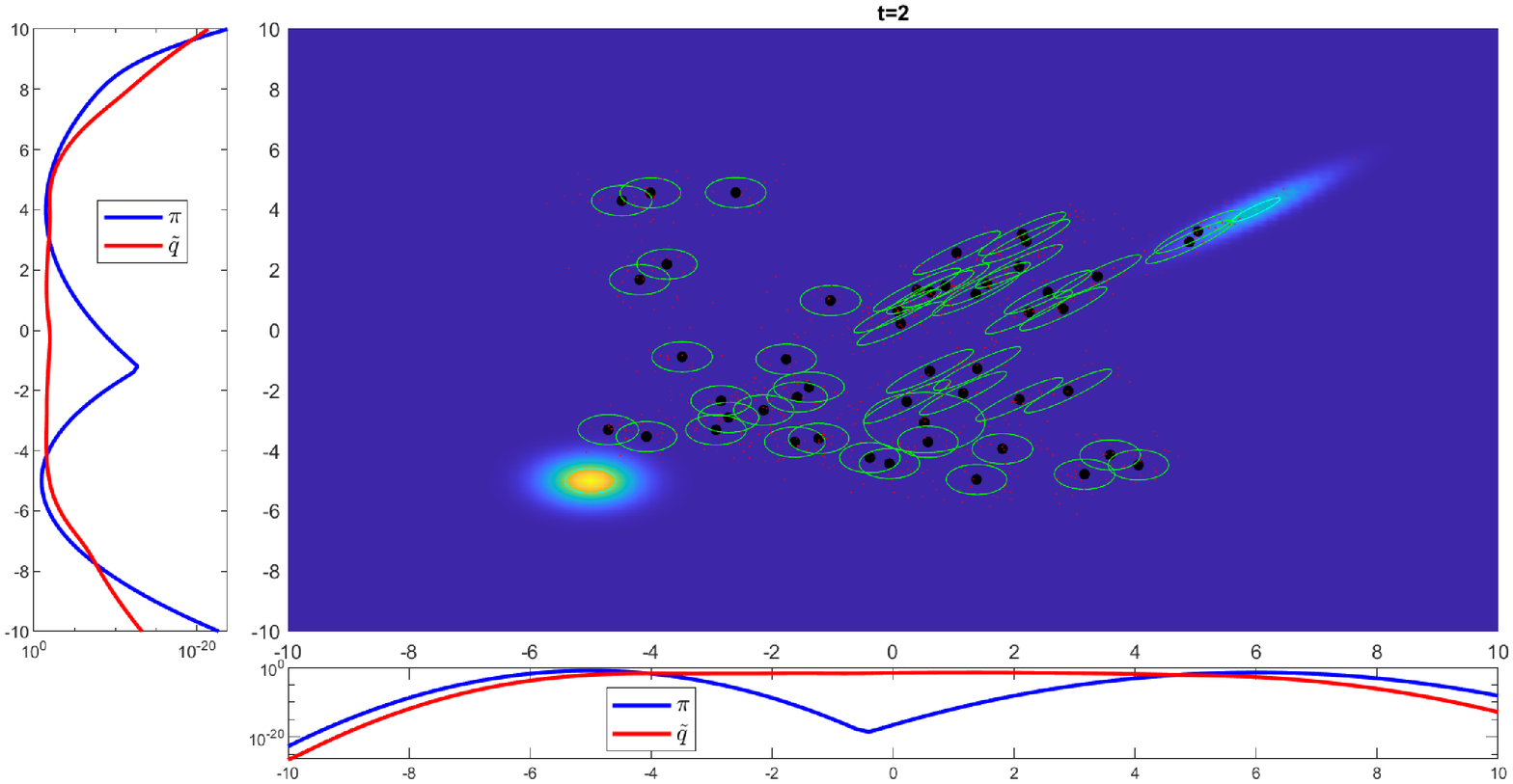} & \includegraphics[width=0.5\columnwidth]{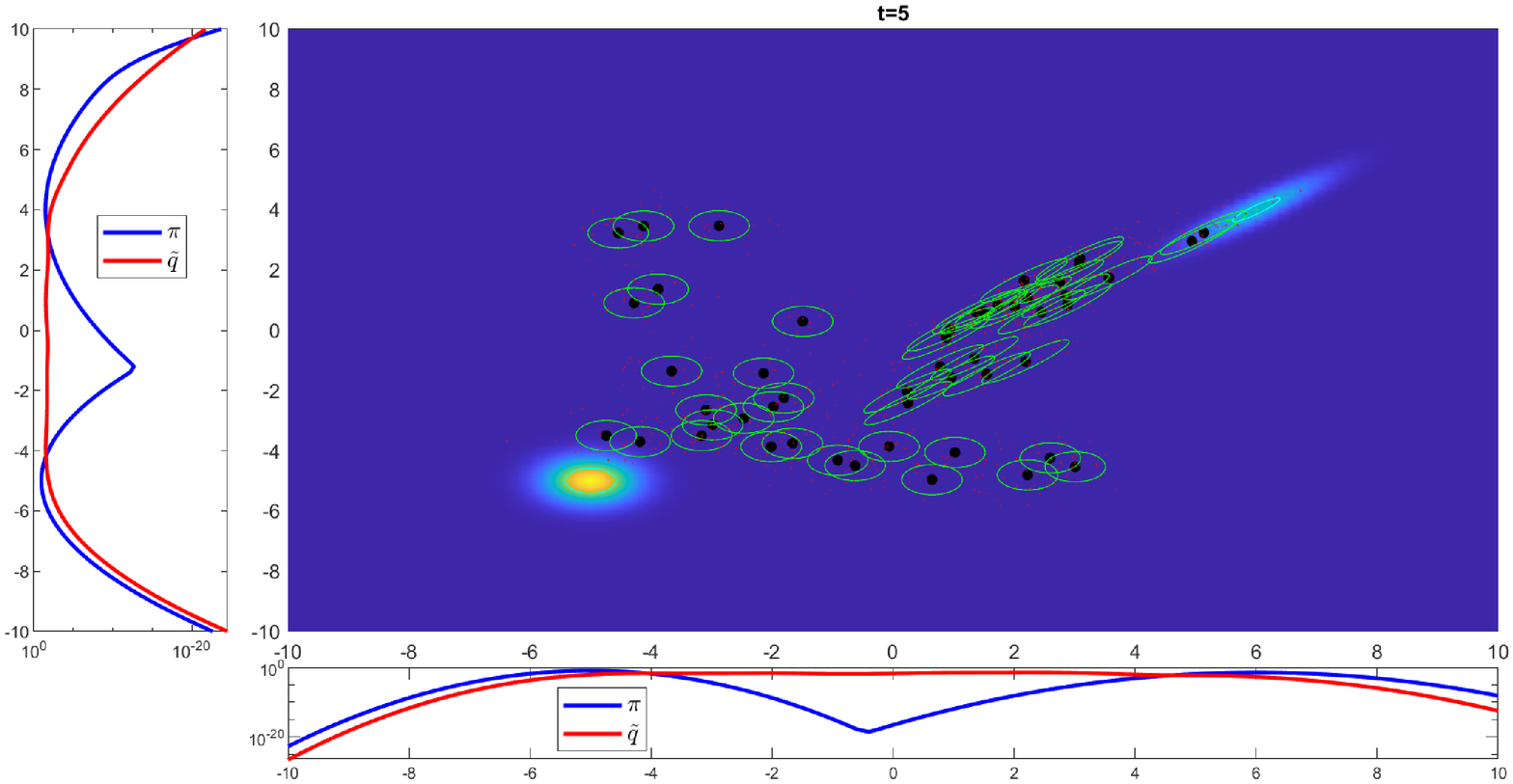} & \includegraphics[width=0.5\columnwidth]{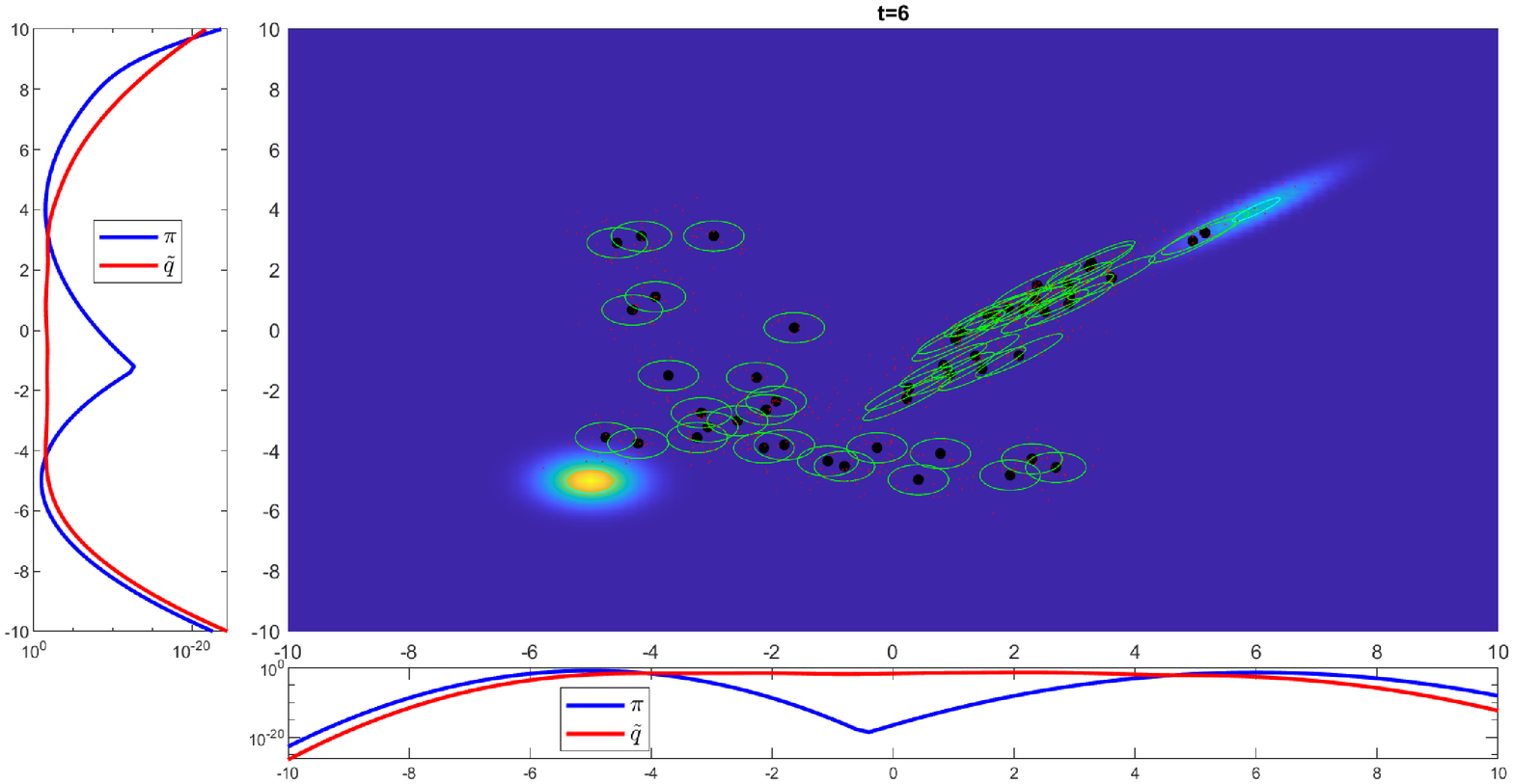} & \includegraphics[width=0.5\columnwidth]{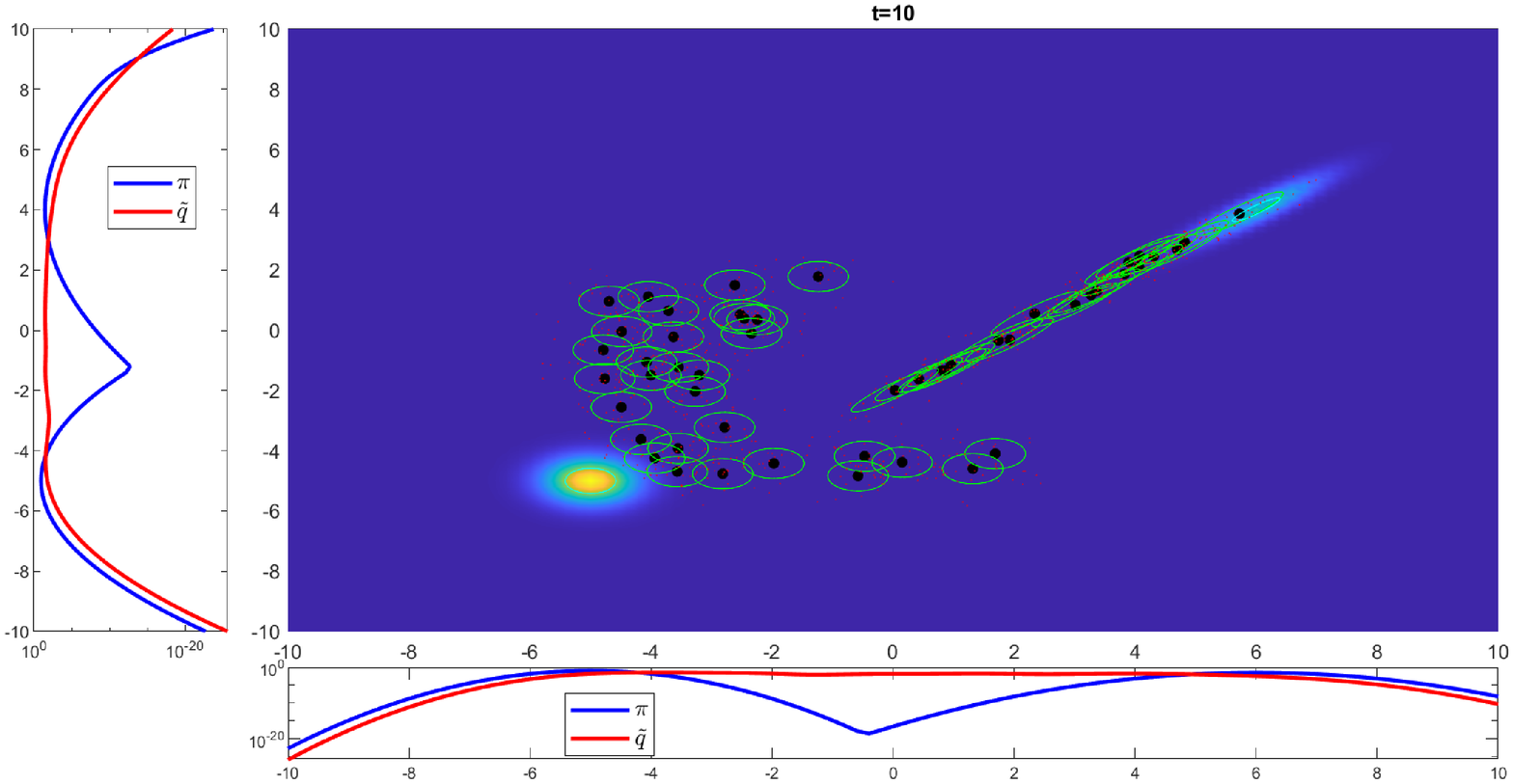}\\
\cblue{GAPIS ($t=2$)} & \cblue{GAPIS ($t=5$)} & \cblue{GAPIS ($t=6$)} & \cblue{GAPIS ($t=10$)}\\
\includegraphics[width=0.5\columnwidth]{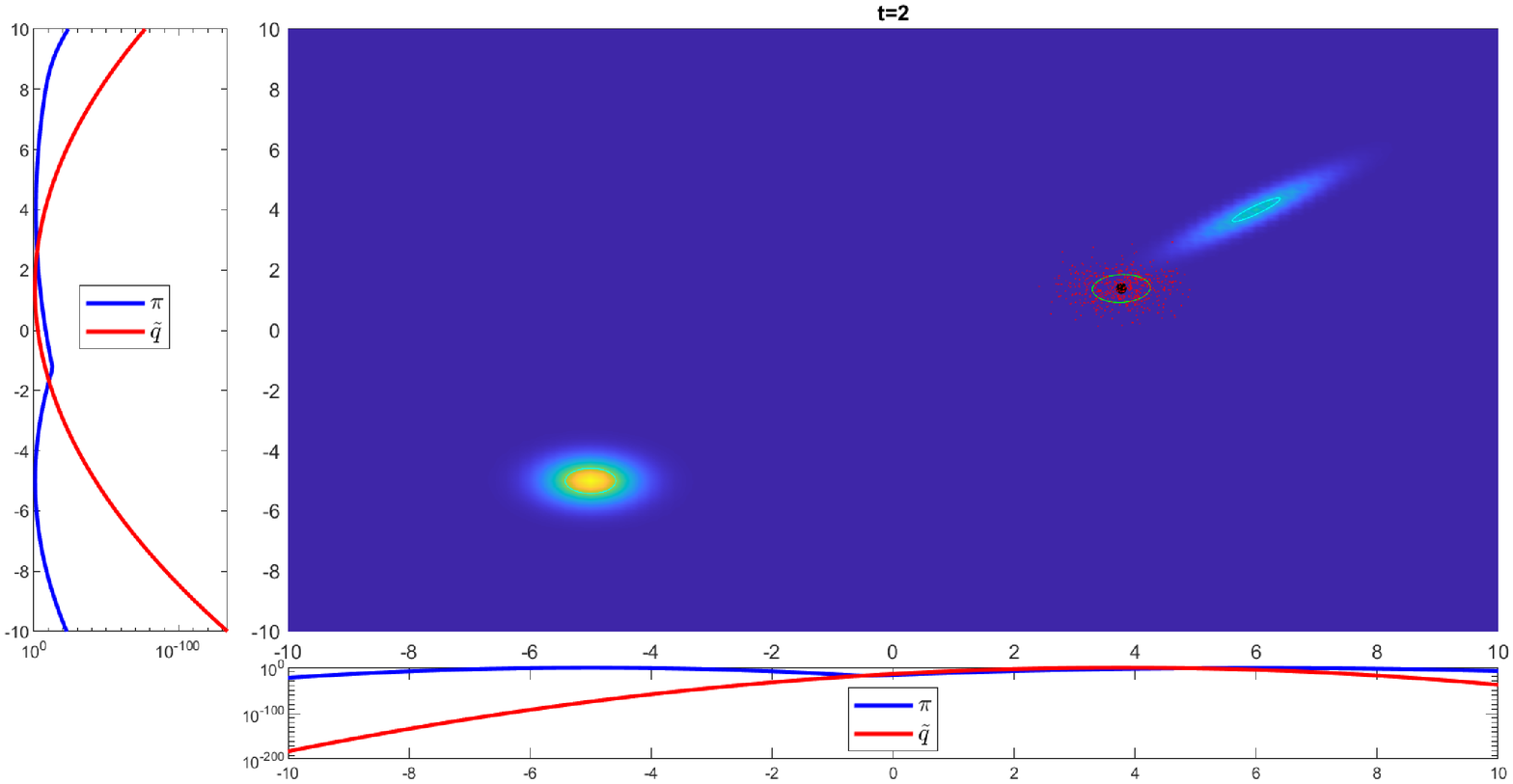} & \includegraphics[width=0.5\columnwidth]{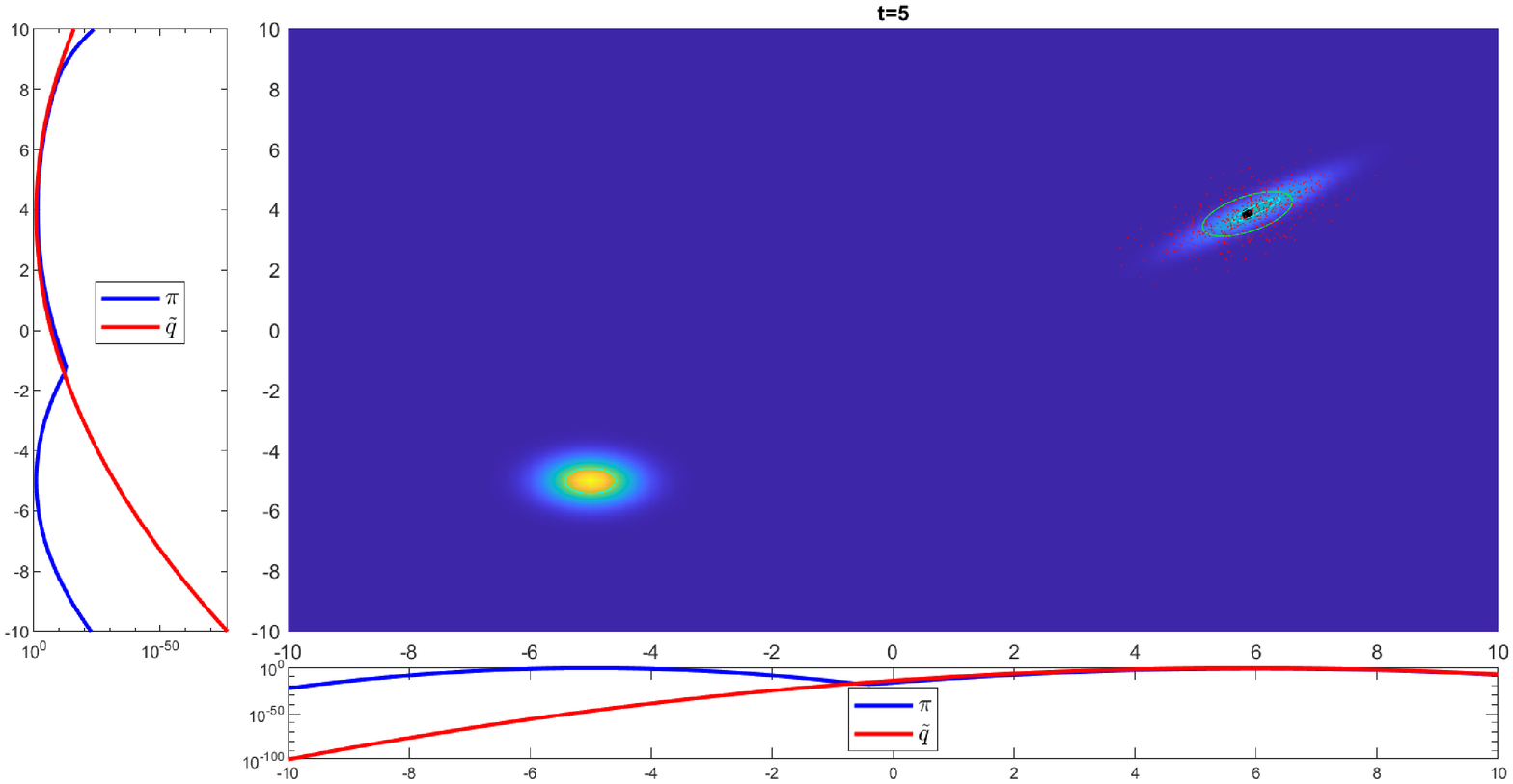} & \includegraphics[width=0.5\columnwidth]{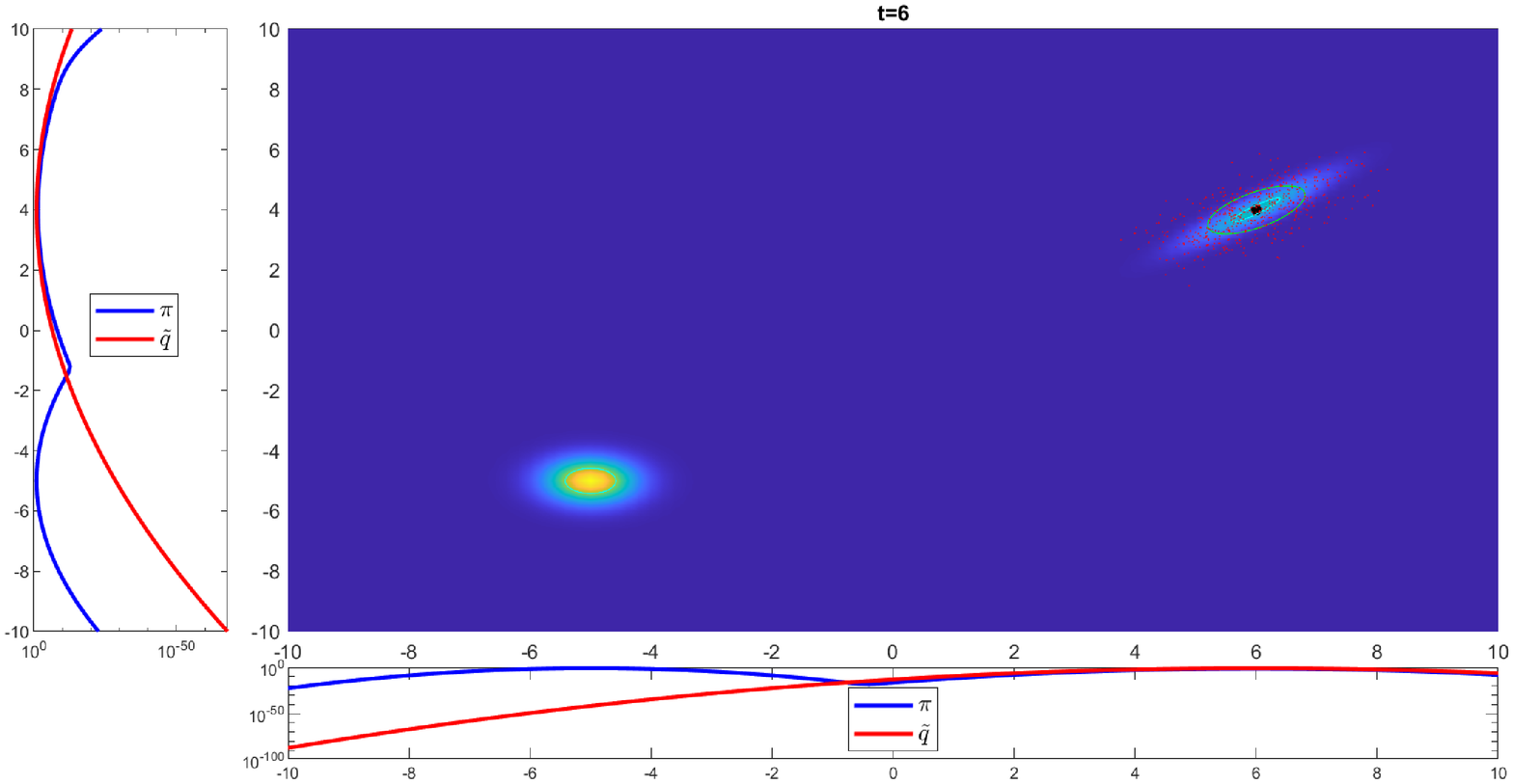} & \includegraphics[width=0.5\columnwidth]{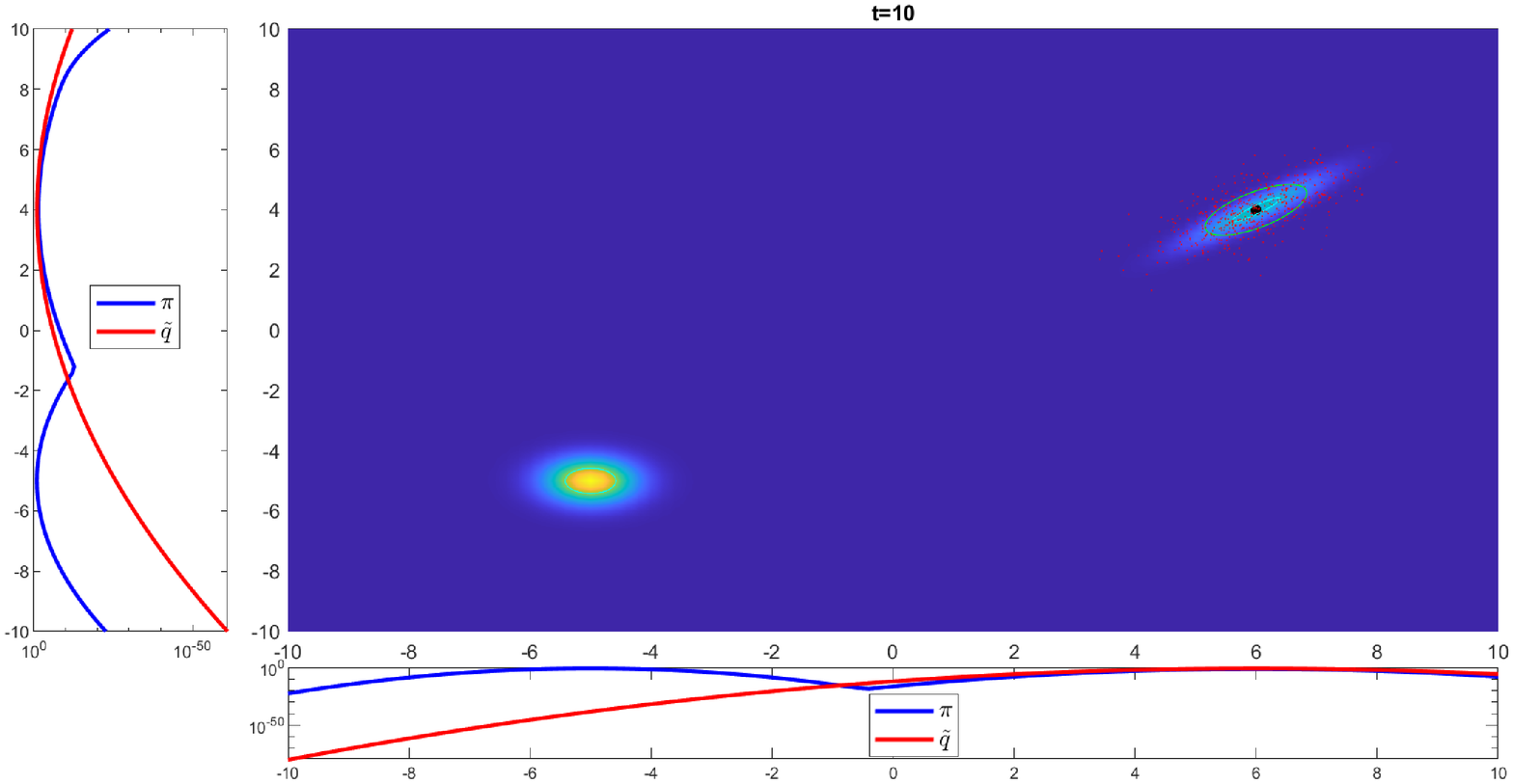}\\
\cblue{AMIS ($t=2$)} & \cblue{AMIS ($t=5$)} & \cblue{AMIS ($t=6$)} & \cblue{AMIS ($t=10$)}\\
\includegraphics[width=0.5\columnwidth]{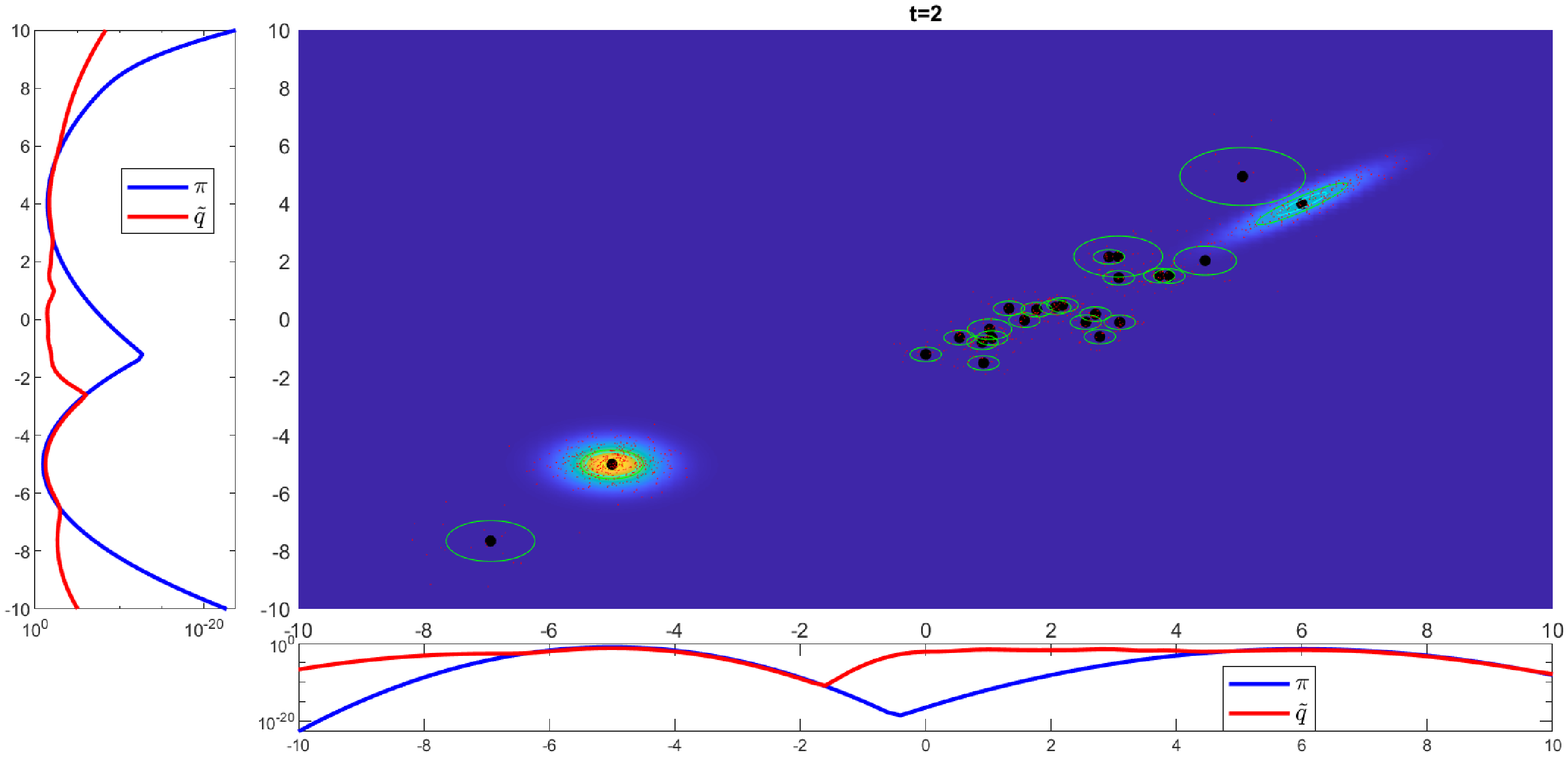} & \includegraphics[width=0.5\columnwidth]{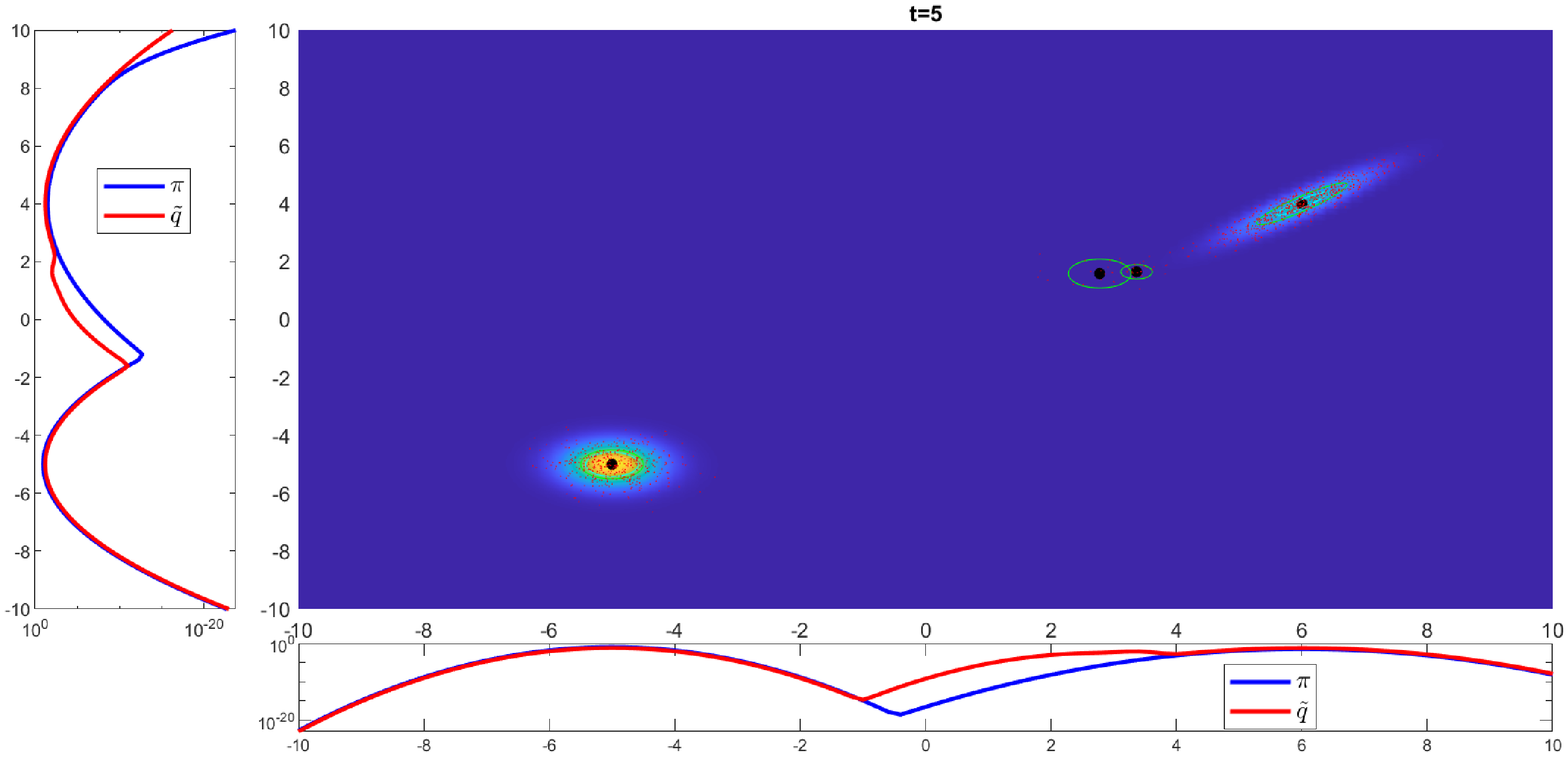} & \includegraphics[width=0.5\columnwidth]{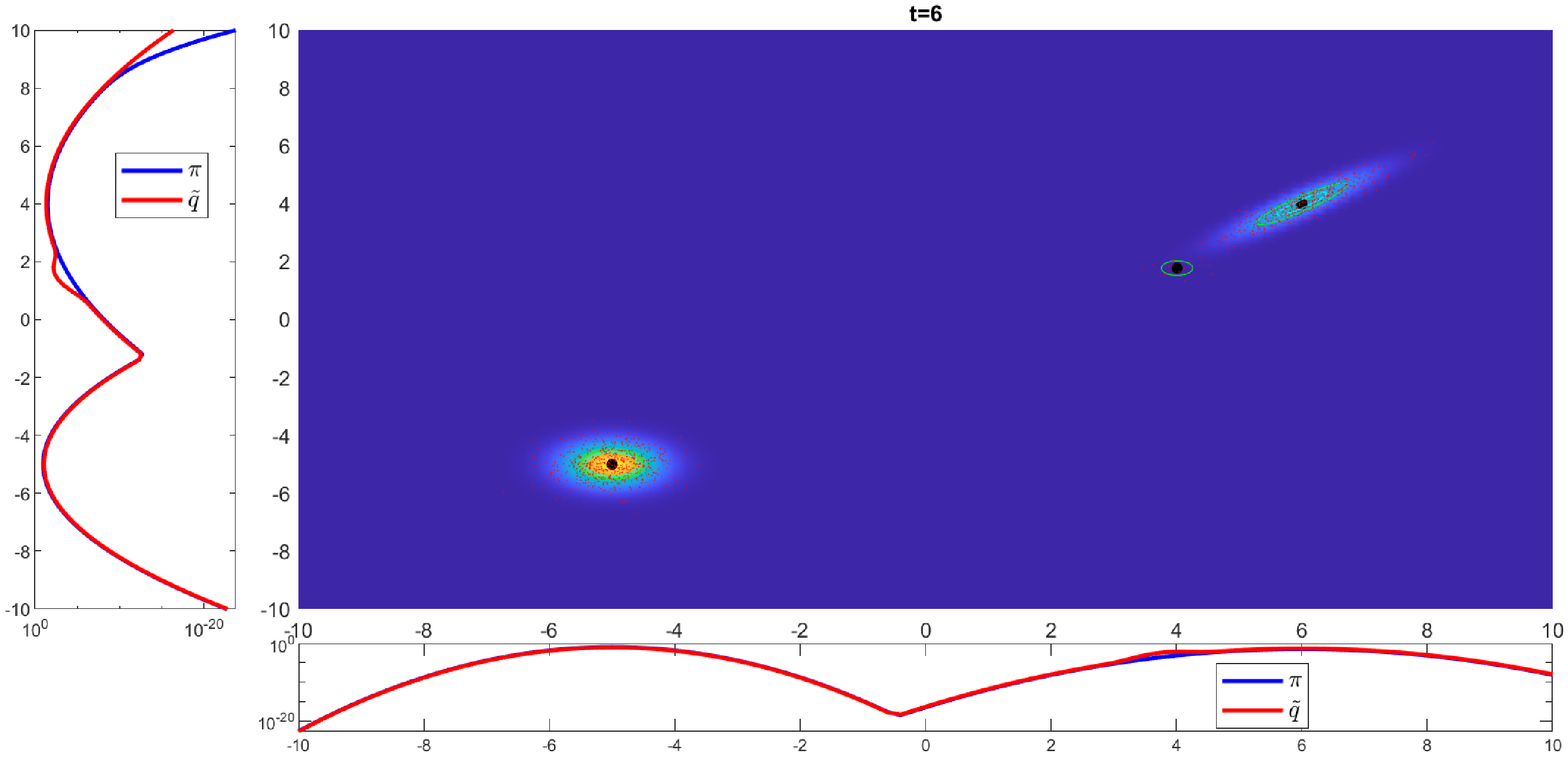} & \includegraphics[width=0.5\columnwidth]{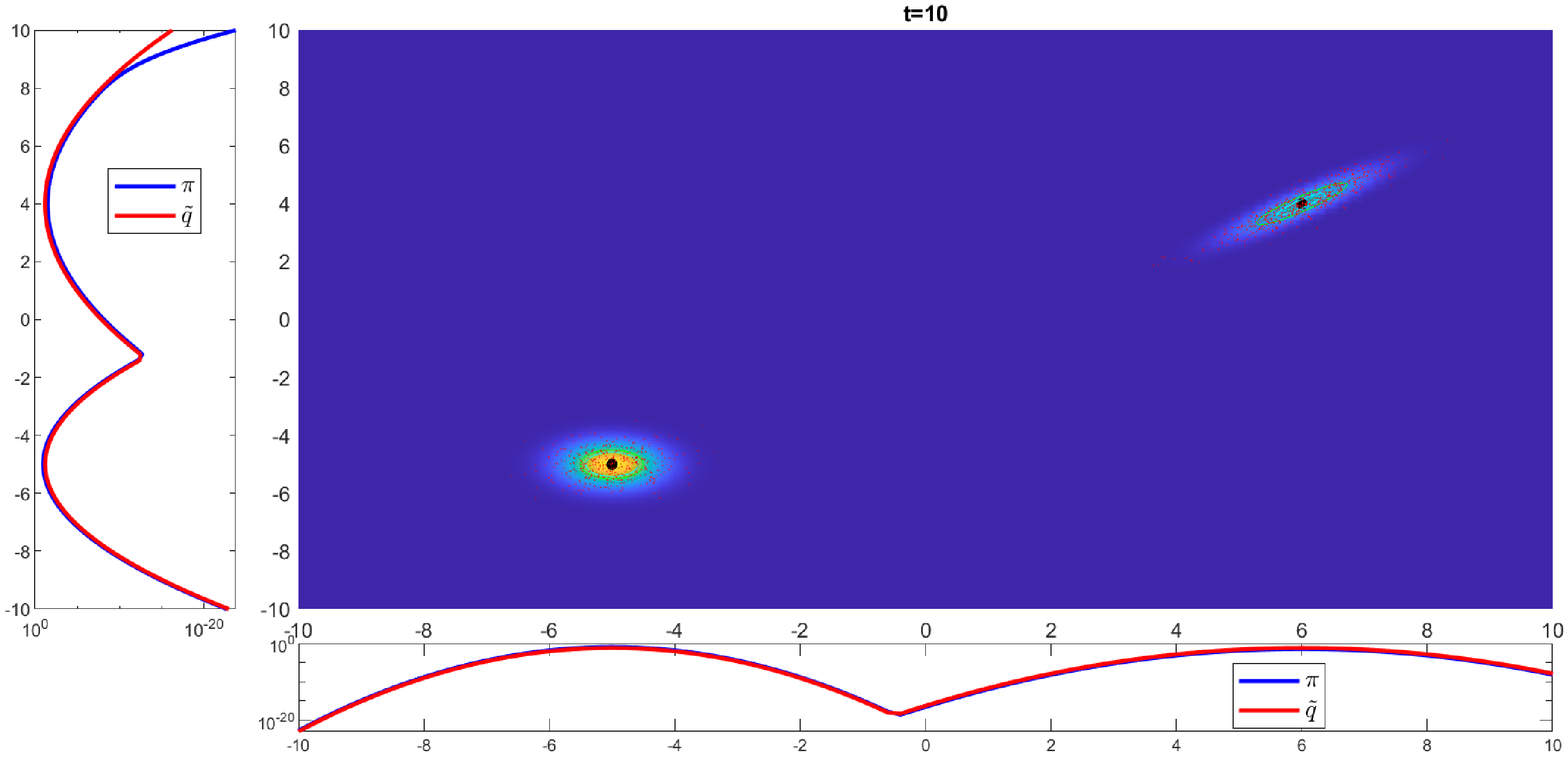}\\
\acro \ - LR ($t=2$) & \acro \  - LR ($t=5$) & \acro \ - LR  ($t=6$) & \acro \ - LR  ($t=10$)\\
\includegraphics[width=0.5\columnwidth]{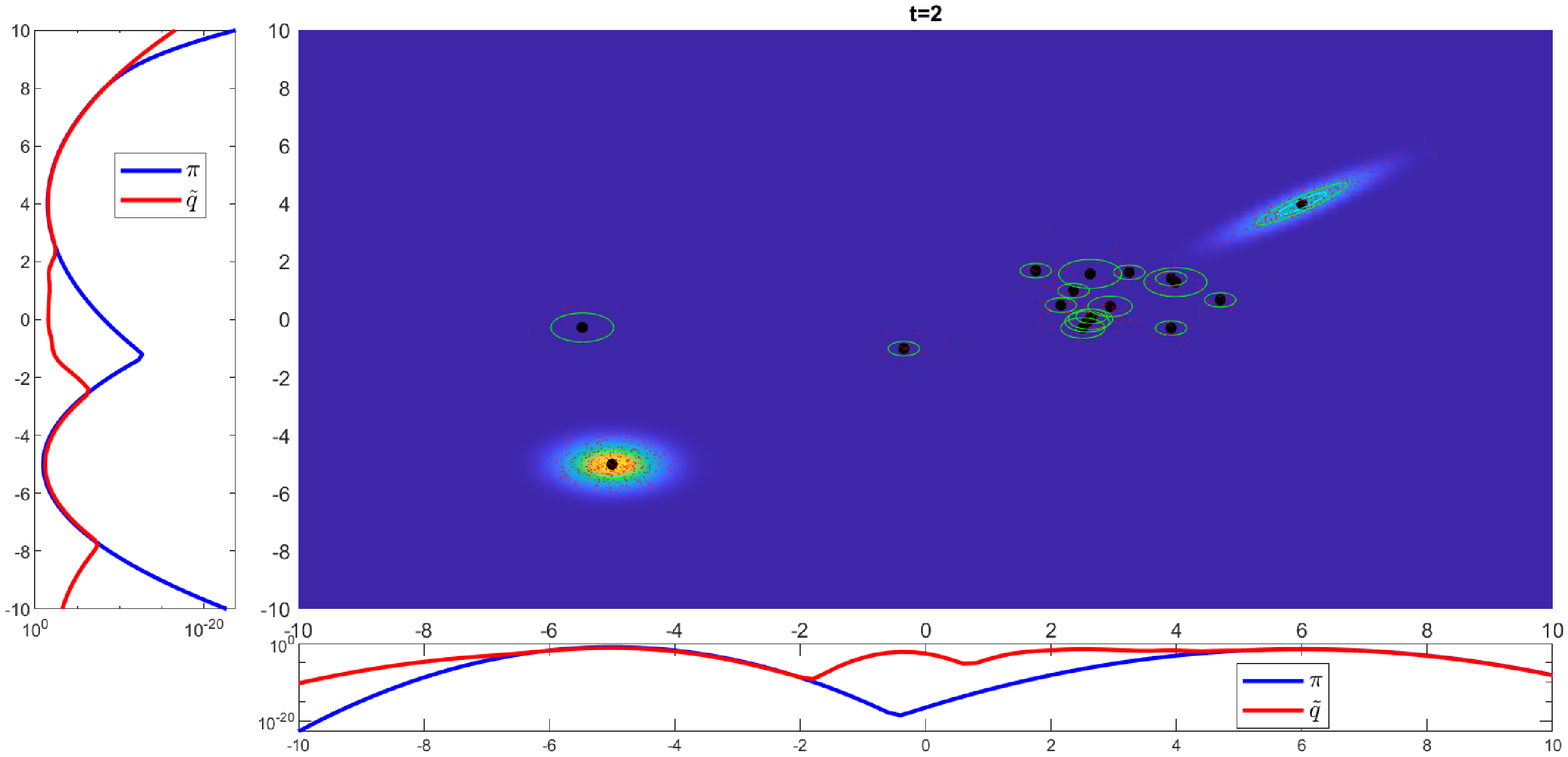} & \includegraphics[width=0.5\columnwidth]{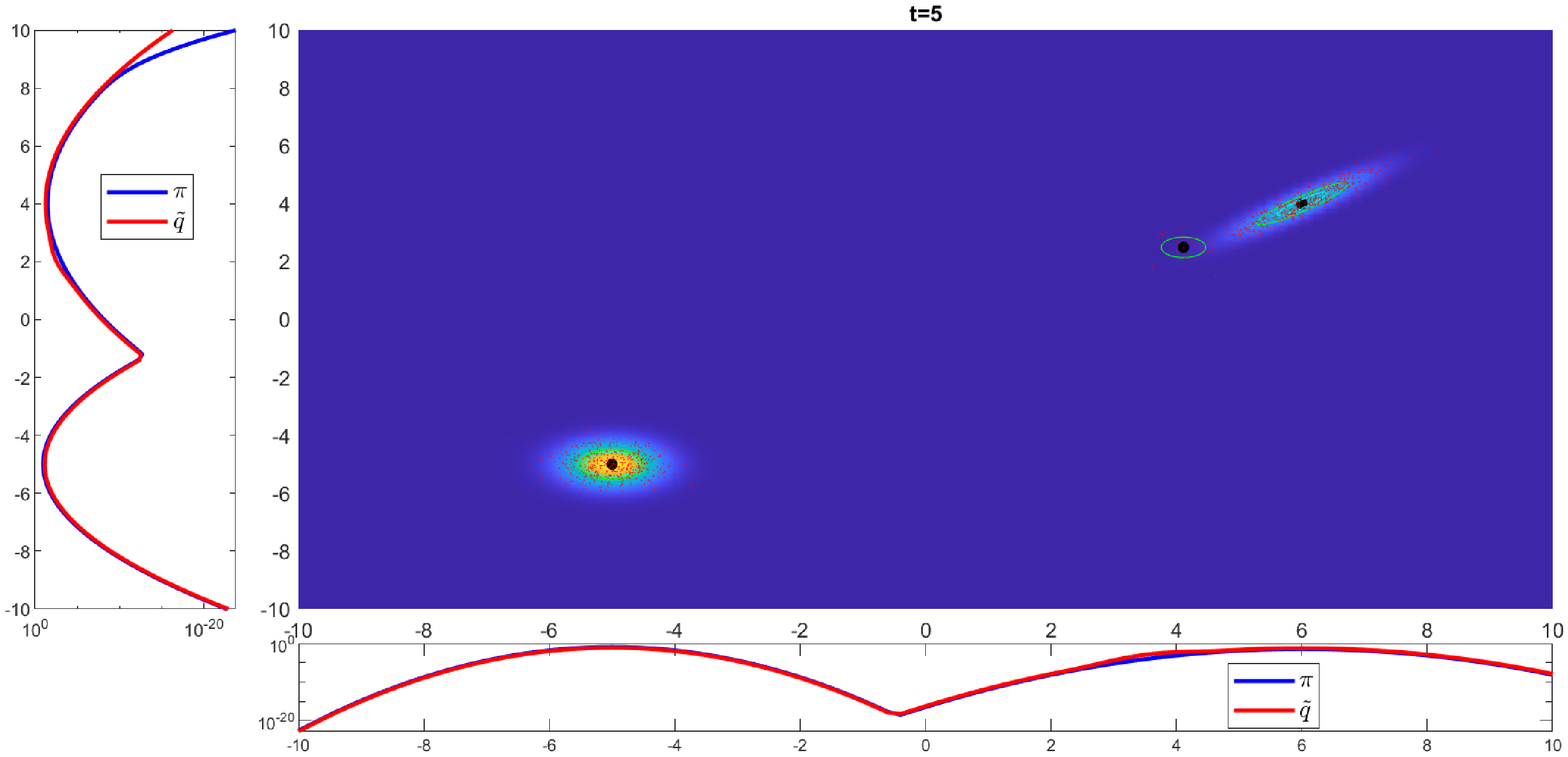} & \includegraphics[width=0.5\columnwidth]{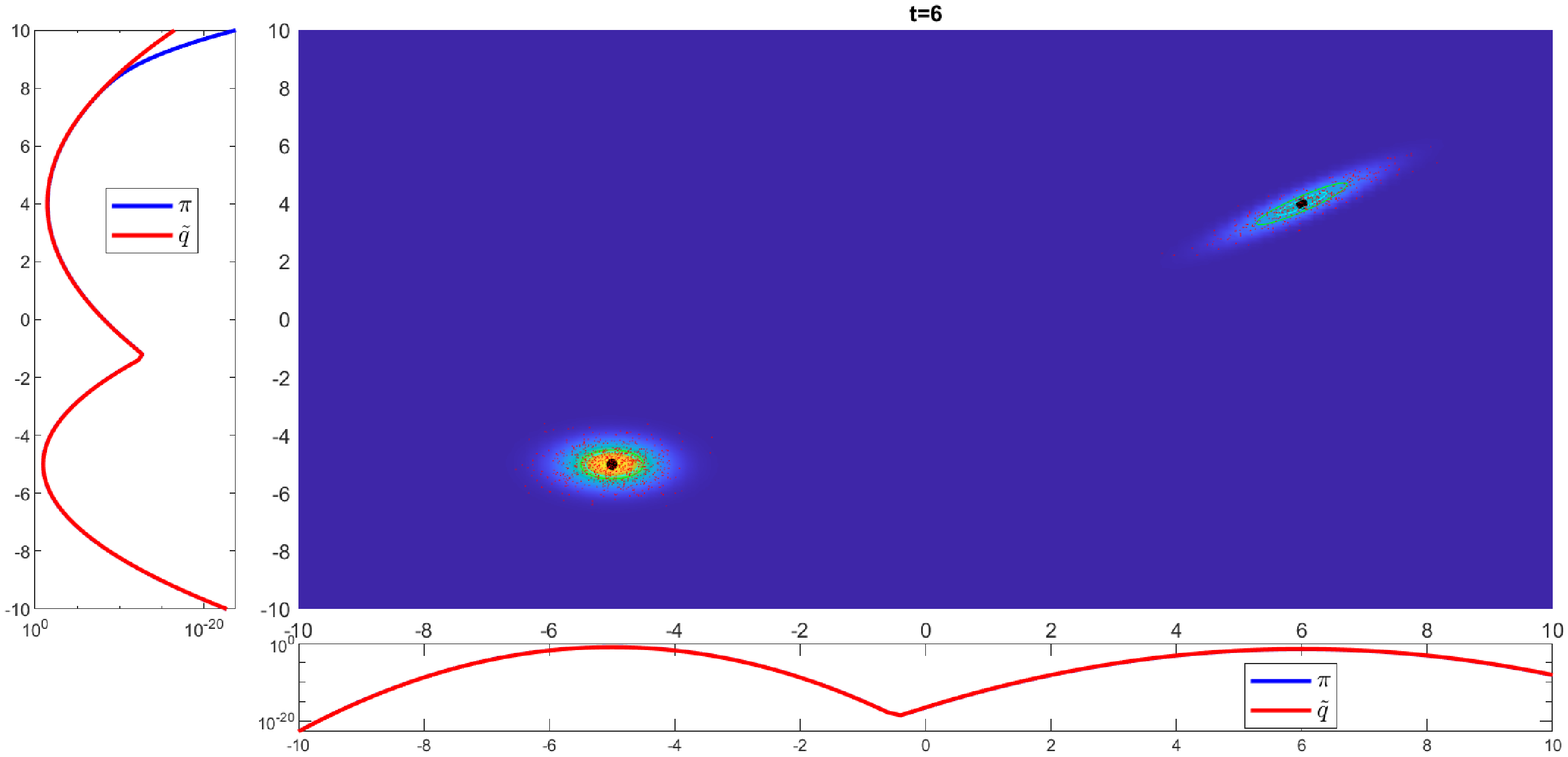} & \includegraphics[width=0.5\columnwidth]{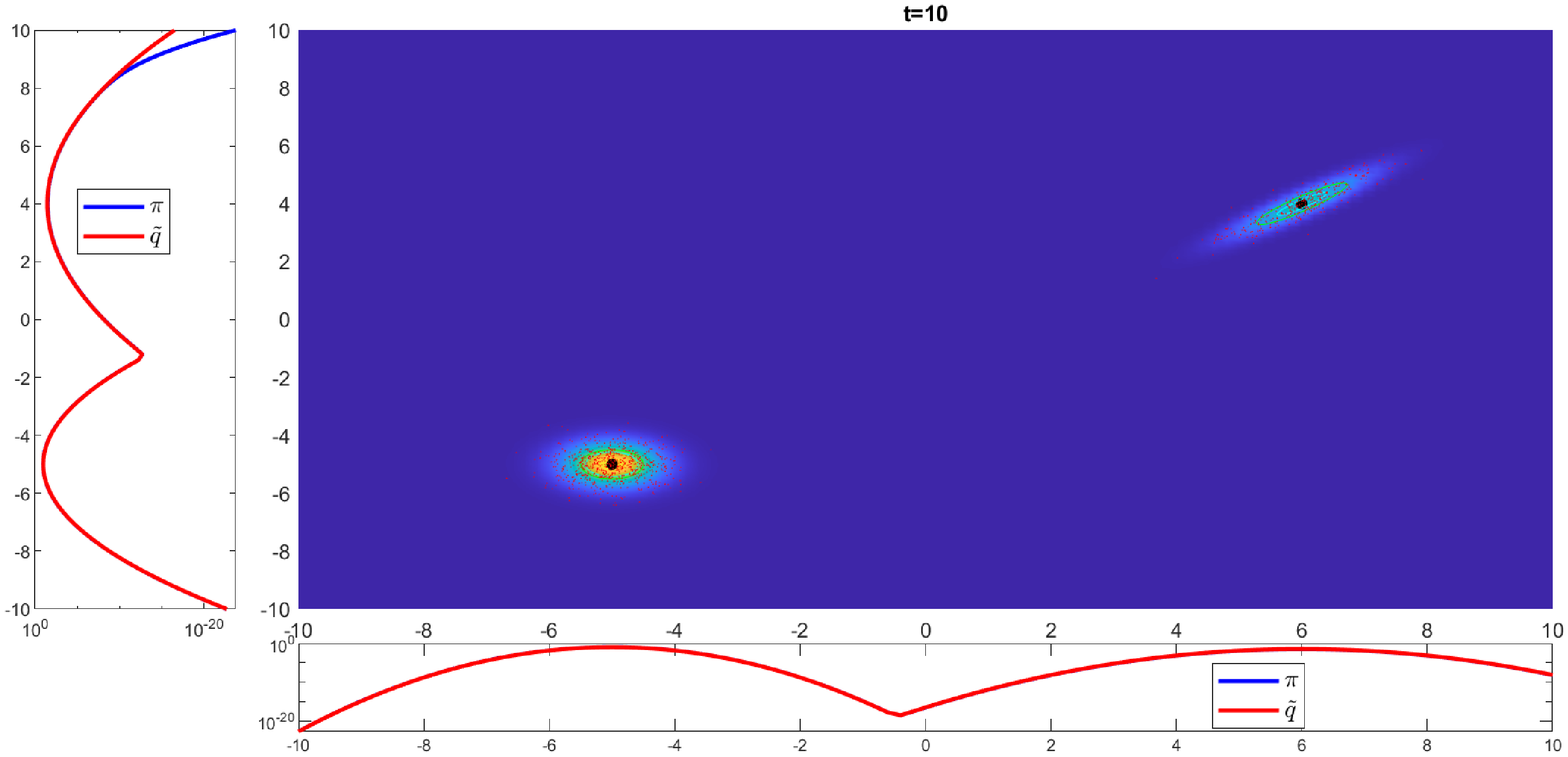}\\
\acro \ - GLR ($t=2$) & \acro \  - GLR ($t=5$) & \acro \ - GLR  ($t=6$) & \acro \ - GLR  ($t=10$)\\
\end{tabular}
\caption{\textbf{Toy example.} Evolution of the reconstructed target along iterations for LR-PMC, GR-PMC, \cblue{GAPIS, AMIS,} and \acro \ for both LR and GLR (with $\Delta = 5$). One can notice the fast convergence of the proposed \acro. The great impact of GLR can be seen, by comparing both \acro \ variants (LR / GLR) between time $t=5$ to time $t=6$, i.e. after and before applying the GR step in the GLR approach.}
\label{fig:toy_example_good_sigma}
\end{figure*} 
\section{Numerical Results}
\label{sec_exp}

In this section, we present several sets of experiments, in order to assess the performance of the proposed \acro~algorithm. Three examples will be considered for the target, namely (i) two-dimensional Gaussian mixture, (ii) multi-dimensional banana-shaped distribution, (iii) posterior distribution arising in a spectral analysis problem. These examples are representative as they incorporate challenging features related to multi-modality and high dimensionality. In all examples, we compare with competitive state-of-the-art adaptive importance sampling techniques, namely GR-PMC and LR-PMC that are two variants of the DM-PMC algorithm \cite{elvira2017improving} {(GR-PMC and LR-PMC)}, AMIS \cite{CORNUET12} and GAPIS \cite{elvira2015gradient}. \cblue{Let us notice that, for the retained settings, the time spent by the methods are actually very similar, which confirms the fairness of our comparisons.}

\subsection{Mixture of Gaussians}
\label{ex_mixture}
%Let us start with our first example, that is the exploration of the mixture of Gaussian distribution. W
In this first example, we consider a multimodal target which is a mixture of five bivariate Gaussian pdfs (i.e. ${d_x} = 2$):
\begin{equation}
(\forall \x \in \mathbb{R}^2) \quad \widetilde{\pi}({\bf x}) = \frac{1}{5}\sum_{i=1}^5 \mathcal{N}({\bf x};{\bf \gamma}_i,\bC_i).
\end{equation}
Here, we set the means ${\bf \gamma}_1=[-10, -10]^{\top}$, ${\bf \gamma}_2=[0, 16]^{\top}$, ${\bf \gamma}_3=[13, 8]^{\top}$, ${\bf \gamma}_4=[-9, 7]^{\top}$, ${\bf \gamma}_5=[14, -4]^{\top}$, and the covariance matrices $\bC_1=[5, \ 2; 2, \ 5]$, $\bC_2=[2, \ -1.3; -1.3, \ 2]$, $\bC_3=[2, \ 0.8; 0.8, \ 2]$, $\bC_4=[3, \ 1.2; 1.2, \ 0.5]$ and $\bC_5=[0.2, \ -0.1; -0.1, \ 0.2]$. The main challenge in this example is  the ability in discovering the $5$ different modes of $\bar{\pi}({\bf x})\propto \pi({\bf x})$. We focus in our tests on the approximation of three quantities, namely the target mean $\E_{\widetilde \pi}[{\bf X}]={[2.4, 3.4]}^{\top}$,  the second moment $\E_{\widetilde \pi}[{\bf X}^2]={[101.04,98.94]}^{\top}$, and the normalizing constant $Z=1$. Since we know the ground truth for these quantities, we can easily assess qualitatively the performance of the different techniques. Furthermore, since the problem is low dimensional, it is possible to approximate the posterior with a very thin grid, allowing to compare visually the performance of the different sampling schemes.

Except for AMIS, in all other algorithms we set $N=50$ proposals (randomly initialized in the square {$[-15,15]\times[-15,15]$}), $T=20$ iterations, and $K=20$ samples per proposal and iteration. Since AMIS has a single proposal, we set $N=1$, $T=500$ and $K=40$, i.e., keeping the same number of target evaluations for a fair comparison. For all algorithms we use {isotropic Gaussian proposals with standard deviation $\sigma\in\{1,3,5\}$, except for \acro, where the proposals are initialized using $\bSigma_n^{(1)} = \sigma^2\mathbf{I}_2$, with $\sigma=5$ and then adapted over the iterations}. In the GLR version of \acro, we set the period $\Delta=5$. %\victor{just add the period of the GLOCAL equal to 5. We need to define the GLOCAL and its notation}. 
In Table \ref{table_exp1} we display the relative mean square error (RMSE) of the AIS estimators. We build the estimators by averaging all the weighted samples of the second half of the iterations, which allows to better determine the adaptive capabilities of each algorithm. We see that the novel \acro~outperforms all other algorithms, in most cases by several orders of magnitude.
 
\begin{table*} [h!]
\scriptsize{
\setlength{\tabcolsep}{2pt}
\def\marginwidth{1.5mm}
\begin{center}
\begin{tabular}{|c||c|c|c||c|c|c||c|c|c||c|c|c||c|c|}                                 
\hline 
&  \multicolumn{3}{c }{ GR-PMC} &  \multicolumn{3}{|c|}{ LR-PMC} &  \multicolumn{3}{|c|}{GAPIS}  & \multicolumn{3}{|c|}{AMIS} & \multicolumn{2}{||c|}{ \acro} \\
\cline{2-15}
&  $\sigma=1$  & $\sigma=3$   & $\sigma=5$  &  $\sigma=1$  & $\sigma=3$   & $\sigma=5$  & $\sigma=1$  & $\sigma=3$   & $\sigma=5$ & $\sigma=1$  & $\sigma=3$   & $\sigma=5$ & LR & GLR\\
\hline
\hline
$Z$ &  0.0432 &0.0066 &0.0073 & 0.0025 &0.0031 &0.0161 & 0.4882 &0.0409 &0.0481 & 0.9836 &0.9814 &0.9487 & \textbf{4 $\cdot 10^{-4}$} & \textbf{4 $\cdot 10^{-4}$}\\
\hline
$\E_{\widetilde \pi}[{\bf X}]$ & 2.4280 &0.4846 &0.3599 & 0.2229 &0.2291 &0.6367 & 2.5397 &1.7318 &1.2595 & 54.5381 &51.0631 &23.4267 & \textbf{0.03532} & 0.03583 \\
\hline
$\E_{\widetilde \pi}[{\bf X}^2]$ & 4.4581 &0.4571 &0.5014 & 0.2244 &0.2203 &0.7778 &2.7414 &1.4743 &2.1444 & 31.9803 &30.1377 &21.4783 &  \textbf{0.0426} & 0.0434 \\
\hline
% \hline
% \hline
%    \multirow{ 3}{*}{{GR-PMC}}   &   $K=5$   &  0.00022  &    0.0219   &  0.0235  \\ 
% \cline{2-5}
%  &  $K=10$     & 0.00008    &  0.0234     &   0.0246  \\ 
% \cline{2-5}
%    &   $K=50$    &    0.00006  &    0.0173   &   0.0517  \\ 
% \hline
% \hline
% \multirow{ 3}{*}{{LR-PMC}}    &   $K=5$   &  \textbf{0.00004}   &    0.0008  &  0.0265  \\ 
% \cline{2-5}
%  &  $K=10$     &  0.00006   &  \textbf{0.0003}    &  0.0238   \\ 
% \cline{2-5}
%   &   $K=50$    &  0.00009   &   0.0009   & \textbf{ 0.0183 } \\ 
% \hline
% \hline
% \multirow{ 3}{*}{{PR-PMC}}    &   $K=5$   &  0.00116   &  0.0040     & {0.0248}   \\ 
% \cline{2-5}
%    &  $K=10$     & \textbf{0.00004 }  &   0.0035   & {0.0234}  \\ 
% \cline{2-5}
%   &   $K=50$    &  0.00061   &  0.0018    & {0.0295}  \\     
% \hline                                                         
\end{tabular}
\end{center}
\caption{\cblue{\textbf{Example \ref{ex_mixture}.} Relative MSE in the estimation of $Z$, $\E_{\widetilde \pi}[{\bf X}]$, and $\E_{\widetilde \pi}[{\bf X}^2]$ in GM2D example. For \acro, we set the initial proposal variance to $\sigma = 5$. The period for GLR is set to $\Delta = 5$. In all PMC-based methods, $(N,K,T) = (50,20,20)$ while $(N,K,T) = (1,500,40)$ for AMIS.} %\emilie{I display results for \eqref{eq:covadapt2}/\eqref{eq:covadapt1a}-\eqref{eq:covadapt1b}.} 
%$ap.past-hessian = 0/ap.past-hessian = 1$}.
\label{table_exp1}
}
}
\end{table*}

\subsection{High-dimensional banana-shaped distribution}
 \label{ex_banana}
The second example focuses on the banana-shaped distribution \cite{haario1999adaptive,Haario2001}. This target shape has been widely used in the past for assessing the performance of sampling methods, as it is particularly challenging to approximate precisely, especially when the dimension of the problem increases. Let us consider a $d_x$-dimensional multivariate Gaussian r.v. $\bar{\X} \sim \mathcal{N}(\x;\textbf{0}_{d_x},\bC)$ with $\bC = \text{diag}(c^2,1,...,1)$. The banana-shaped distribution is defined 
as the pdf of the transformed multivariate variable $(X_j)_{1 \leq j \leq d_x}$ such that $X_j = \bar X_j$ for $j\in \{1,...,d_x\}\setminus 2$, and $X_2 = \bar X_2 - b(\bar X_1^2 - c^2)$. Hereabove, $b$ and $c$ are shape parameters set in the sequel to be equal to $c=1$ and $b=3$. {We evaluate the performance of different AIS methods in estimating $\E_{\widetilde \pi}[{\bf X}]$, for different dimensions $d_x \in \{2,5,10,15,20,30,40,50 \}$. {All algorithms initialize the location parameters of the proposals randomly and uniformly within the square $[-4,4]\times[-4,4]$, and 1000 independent runs are performed.  } %\emilie{let us not forget the initialization settings}. 
In all algorithms, except AMIS, we set $N=50$, $K=20$, and $T=20$. In AMIS, we set $N=1$, $K=500$ and $T=40$, for a fair comparison in terms of total number of target evaluations (we recall AMIS imposes a unique proposal). In \acro, the initial proposal covariances are isotropic, $\bSigma_n^{(1)} = \sigma^2 \mathbf{I}_{d_x}$, with $\sigma =3$, and we implement the resampling strategies LR and GLR (with $\Delta=5$). The other algorithms are initialized also with isotropic covariances with $\sigma \in \{1, 3, 5\}$. In Table \ref{table_exp2_dim5} we show the MSE of the proposed \acro~and its competitors in the estimation of the target mean for dimensions $d_x\in \{5, 20, 50 \}$. We also display in Fig.~\ref{fig:banana_slpmc} the performance of \acro, LR-PMC and GR-PMC, measured in terms of MSE averaged across dimensions. In this example, the best performance is reached with the LR version of the \acro, {followed by the GLR version of the same algorithm. AMIS is the second best algorithm in most cases, possibly due to the covariance adaptation that it incorporates (unlike the GR-PMC and LR-PMC schemes). The competitors GR-PMC and LR-PMC degrade when the dimension decreases. Note that the MSE of our \acro \; decreases with the dimension. This can be explained by the particular structure of the target, which is conditionally Gaussian in all dimensions except one, and hence it represents a challenge for \acro \ only in that particular dimension}.} %\emilie{analyze the results, for the covariance update rule ? or we keep only one of both ?} \emilie{any reason for poor performance of GLR here?} \victor{unclear to me...}
%We see that \acro \; obtains the best results for all dimensions. In this particular experiment, 

\begin{figure}[htp]
\centering
\includegraphics[width=0.99\columnwidth]{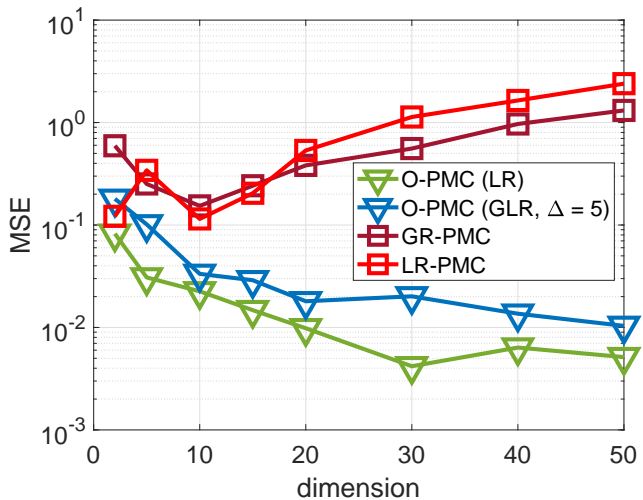}
\caption{\textbf{Example \ref{ex_banana}.} MSE in the estimation of $\E_{\widetilde \pi}[{\bf X}]$ of the banana-shaped distribution versus the dimension $d_x$, with GR-PMC, LR-PMC with $\sigma = 1$ and the proposed \acro~method.%, with GR-PMC (brown square), LR-PMC (red square) with $\sigma = 1$, and the proposed \acro~method with LR (green triangle) or GLR (blue triangle) schemes.
}
\label{fig:banana_slpmc}
\end{figure}

\begin{table*} [h!]
\scriptsize{
\setlength{\tabcolsep}{2pt}
\def\marginwidth{1.5mm}
\begin{center}
\begin{tabular}{|c||c|c|c||c|c|c||c|c|c||c|c|c||c|c|}                                 
\hline 
&  \multicolumn{3}{c }{ GR-PMC} &  \multicolumn{3}{|c|}{ LR-PMC} &  \multicolumn{3}{|c|}{GAPIS}  & \multicolumn{3}{|c|}{AMIS} & \multicolumn{2}{||c|}{ \acro} \\
\cline{2-15}
&  $\sigma=1$  & $\sigma=3$   & $\sigma=5$  &  $\sigma=1$  & $\sigma=3$   & $\sigma=5$  & $\sigma=1$  & $\sigma=3$   & $\sigma=5$ & $\sigma=1$  & $\sigma=3$   & $\sigma=5$ &  LR & GLR\\
\hline
\hline
$d_x = 5$ &  0.2515 &0.1350 &0.2299 & 0.3418 &0.5289 &0.5925 & 0.3007 & 0.3631 & 0.7790& 0.1758 & 0.1783 & 0.1572 
&  \textbf{0.0308}  & 0.1014 \\%back to id / inherit
\hline
$d_x = 20$ &  0.3818 &3.1430 &11.1921 & 0.5340 &6.4936 &23.3693 & 1.5299 & 1.6555 & 1.5640& 0.1901 &  0.1574 & 0.2673 &  \textbf{0.0098}   & 0.0180\\
\hline
$d_x = 50$ &  1.3134 &9.6571 &42.6815 & 2.3963 &21.7097 &6.3350 & 2.5524 & 2.5632 & 2.8486& 0.6074 & 0.7992 & 1.5334 
& \textbf{0.0051}     &  0.0104\\
\hline                                                          
\end{tabular}
\end{center}
\caption{\textbf{Example \ref{ex_banana}.} MSE in the estimation of $\E_{\widetilde \pi}[{\bf X}]$ of the banana-shaped distribution for dimensions $d_x = 5$, $20$ and $50$. For \acro, we set the initial proposal variance to $\sigma = 3$. The period for GLR is set to $5$. In all PMC-based methods, $(N,K,T) = (50,20,20)$ while $(N,K,T) = (1,500,40)$ for AMIS. %GLR with period 5
}
\label{table_exp2_dim5}
}
\end{table*}

%\emilie{add comments on the results. Why performance of \acro~increase with dimension ? maybe the problem becomes more unimodal so optimization-based method helps a lot ?}

%\cred{
%For both variants GR-PMC and LR-PMC, we make use of proposal covariances $\bSigma_n = \sigma^2 \mathbf{I}_{d_x}$, with $\sigma \in \{1,3\}$ (the performance with $\sigma=5$ is much worse and is not displayed). In general, the performance of all methods worsens when the dimension grows. Note that the \acro~outperforms the DM-PMC in all dimensions, keeping a substantial gap in terms of MSE.}

%\victor{Plot MSE evolution w.r.t. iterations for given dimension.}

%\victor{Like Figure 1, each subplot just one sigma, but 3 subplots with different sets of parameters (playing with T/K/N).}

%\victor{representation of 2D evolution of particles and some of the others (4 iterations? 3 methods?, either elipses or reconstruction of}

%\begin{figure}[htp]
%\centering
%\includegraphics[width=0.99\columnwidth]{figs/ex2}
%\caption{\textbf{Example \ref{ex_banana}.} MSE in the estimation of $\E_{\widetilde \pi}[{\bf X}]$ of the banana-shaped distribution versus the dimension $d_x$}
%\label{fig_ex2}
%\end{figure}

\subsection{Spectral analysis}%: estimating the frequencies of a noisy multi-sinusoidal signal
\label{ex:sinus}

Our last example addresses the problem of estimated the parameters of a multi-sinusoidal signal from noisy and undersampled acquisitions of it. We consider the following observation model:
\begin{equation}
(\forall j \in \{1,\ldots,d_y\}) \quad y_j = \sum_{s=1}^S a_s \sin(2\pi \omega_s \tau_j + \varphi_s) + n_j,
\end{equation}
where $(\tau_j)_{1 \leq j \leq d_y}$ defines a discrete uniform time grid, $(n_j)_{1 \leq j \leq d_y}$ a noise assumed to be i.i.d. Gaussian with known variance $\sigma_n^2$, and $(a_s,\omega_s,\varphi_s)_{1\leq s\leq S}$ the amplitude, frequency and phase parameters, respectively, of $S$ sinusoidal components. We focus on the problem of identifying the unknown frequencies and amplitudes, i.e. $d_x = 2 S$, and for all $i \in \{1 \leq i \leq d_x\}$, $x_i = \omega_s$ and $x_{i+S} = a_s$. 

Given the Gaussian model on the noise, the posterior distribution of $\x$ given $\y$ reads $\pi(\x) \propto \exp(-f(\x))$ with 
\begin{equation}
f(\x) = \frac{1}{2 \sigma_n^2} \sum_{j=1}^J \left(y_j - \sum_{s=1}^S a_s \sin(2\pi \omega_s \tau_j + \varphi_s)\right)^2  - \log (p_0(\x)),
\end{equation} 
with $p_0$ the prior distribution on $\x$. {This prior factorizes as $p_0(\x) = p_{\omega}(\x_{1:S})p_{a}(\x_{S+1:2S})$, where $\x_{1:S}$ contains the first $S$ dimensions of $\x$ (i.e., corresponding to the $S$ unknown frequencies), and $\x_{S+1:2S}$ corresponds to the $S$ unknown amplitudes. The prior $p_{\omega}$ is uniform in the support $\{ \x_{1:S}: 0\leq\x_1\leq\x_2...\leq\x_S\leq0.5\}$, i.e., we restrict the frequencies to be defined in increasing order. The prior $p_{a}$ factorizes across all dimensions and is a uniform distribution in {$[0,+\infty[$}. The data is generated by simulating $d_y = 30 \, S$ points {regularly spaced over $[1,d_y]$}. We explore the case with $S \in \{2,3,4,5\}$ (i.e., $d_x \in \{4,6,8,10\}$). We set the observation noise variance to $\sigma_n^2 = 0.5^2$, and the phases $\varphi_s=0$, for $s=1,...,S$.} %Function $f$ is highly multimodal, which makes its exploration challenging, even in low dimensional case. 

{All algorithms simulate the initial location parameters as the prior $p_0$, and the initial covariance matrices are chosen to be isotropic with  $\sigma \in \{ 10^{-3},  10^{-2},  10^{-1}\}$, except \acro \;, where the initialization is done only with $\sigma = 10^{-2}$ since the covariance is adapted. Table \ref{table_exp3_median} shows the median squared error (medianSE) in the estimation of the target mean, considering as ground truth the true frequencies and amplitudes that we have set to generate the data. Note that the target mean can be significantly different from those parameters, and for this reason, we also displayed in Table \ref{table_exp3} the averaged MSE between the signal reconstructed with the estimated parameters w.r.t. the noiseless sequence generated with the true parameters. We observe that with both figures of merit, \acro \; obtains the best results for most setups. While all methods obtain reasonable results for $S=2$ ($d_x=4$), their performance degrade faster than in \acro \; when the dimension of the problem is increased.}

{In Fig. \ref{fig_freqs_convergence}, we display the evolution with the number of iterations of the medianSE in the target mean estimator for GR-PMC, LR-PMC, and \acro \; (LR, GLR with $\Delta=2$, and GLR with $\Delta=5$) algorithms. At each iteration $t$, we compute the estimator with all simulated samples from the beginning, re-normalizing the importance weights to build a unique estimator as it is done, for instance, in \cite{APIS14}. We use the same parameters as those of Tables \ref{table_exp3_median} and \ref{table_exp3}, setting $\sigma = 10^{-2}$. We observe that all algorithms improve when the number of iterations grows, and that all versions of our proposed \acro \; algorithm adapt faster than the competitors. We observe that the GLR version of \acro \; with $\Delta = 5$ adapts faster than the case with $\Delta = 2$, while the best adaptation for this particular setup is obtained by the LR version of \acro. 
}

{Finally, Fig. \ref{fig_freqs_comparison_gt} displays the ground truth and the estimators obtained by GR-PMC and LR-PMC (left subplots) and the LR and GLR versions of \acro (right subplots). We explore the cases with $S=2$ ($d_x = 4$; top subplot), $S=3$ ($d_x = 6$; middle subplot), and $S=4$ ($d_x = 8$; bottom subplot). {The vertical bars represent {the median estimate} $\pm$ the mean absolute deviation (MAD)}. We observe that in all dimensions, the \acro \; obtains closer estimates to the ground truth,  both in the frequencies and in the amplitudes. We note that when the dimension is increased (bottom subplots), the problem becomes more challenging but \acro \; still performs successfully unlike GR-PMC and LR-PMC. 
}

%\victor{You wrote: Bidimensional examples of $f(\x)$ are depicted in Figure. For the moment we don't have these plots. I do not think amplitude plus frequency is as nice as two frequencies (but not sure). Also you wrote ``We also present visual results to assess the exploration performance of our method.'', we dont have it, right? (we have enough I'd say).}\emilie{i agree, and i'd say we have enough too}

%\emilie{let us not forget the initialization settings}. 

%\cred{Bidimensional examples of $f(\x)$ are depicted in Figure ... We consider a uniform prior to constrain the frequencies to be sorted and to belong to the set $[0,\frac{1}{2}]^S$. Positivity constraint was furthermore added on the amplitude values.}

 %\cred{We apply \acro, ...., with settings ... The results are evaluated either by mean of the MSE between the mean estimator and the true parameters \cred{or with the reconstructed MSE ...}. }

%\subsubsection{Unknown frequencies}
%\victor{only freqs and amplitudes right?}
%\subsubsection{Unknown frequencies and amplitudes}

\begin{figure}[htp]
\centering 
\includegraphics[width=1.1\columnwidth]{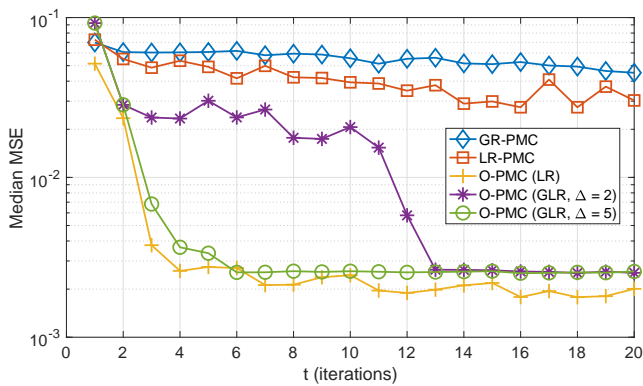}
\caption{{\textbf{Example \ref{ex:sinus}.} {Evolution of the medianSE with respect to ground truth amplitudes and frequencies for dimension $d_x=4$ as function of the number of iterations of GR-PMC, LR-PMC, and \acro.}}}
\label{fig_freqs_convergence}
\end{figure}

\begin{figure}[htp]
\centering
\begin{tabular}{@{}c@{}c@{}}
\includegraphics[width=0.49\columnwidth]{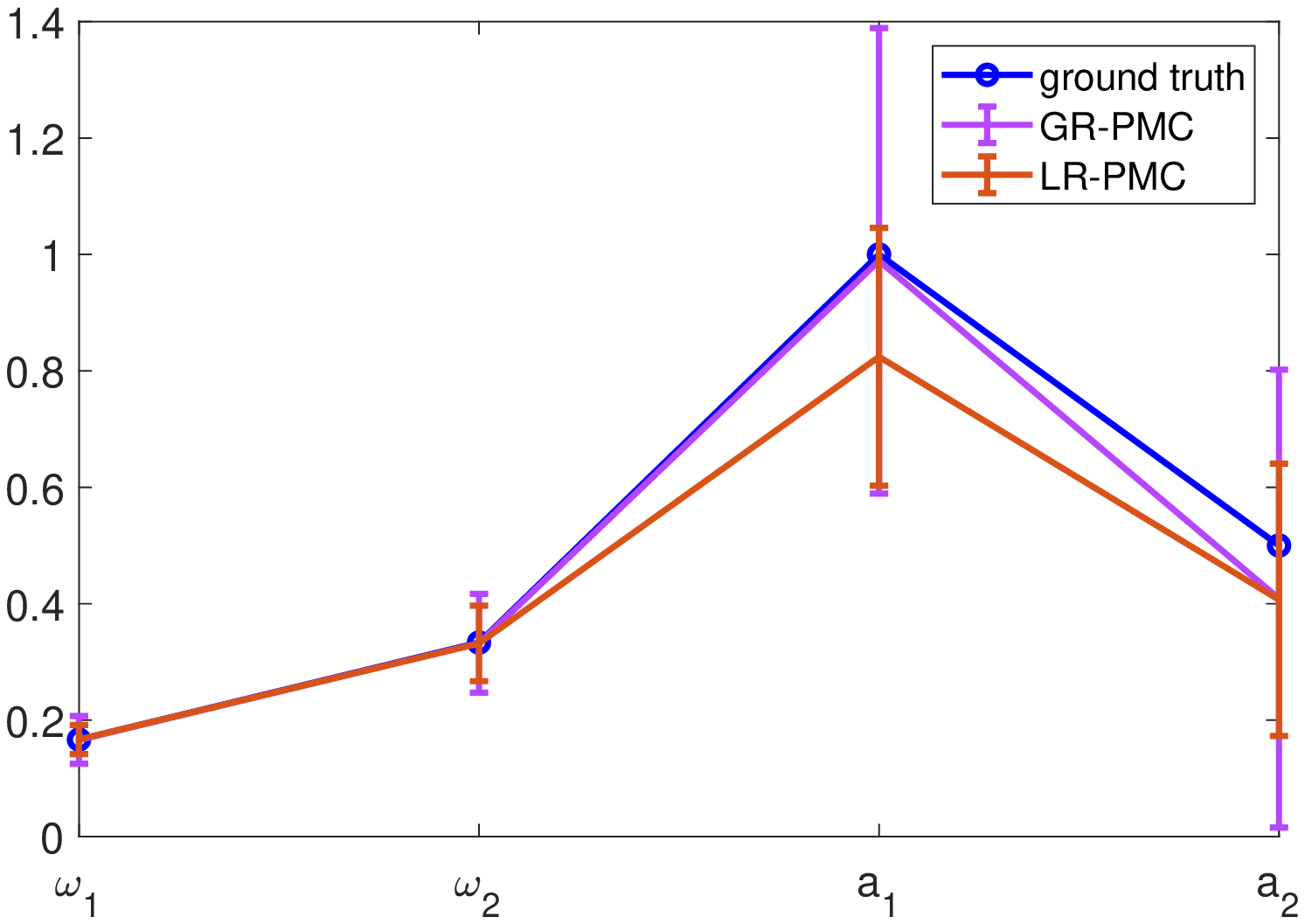} & \includegraphics[width=0.49\columnwidth]{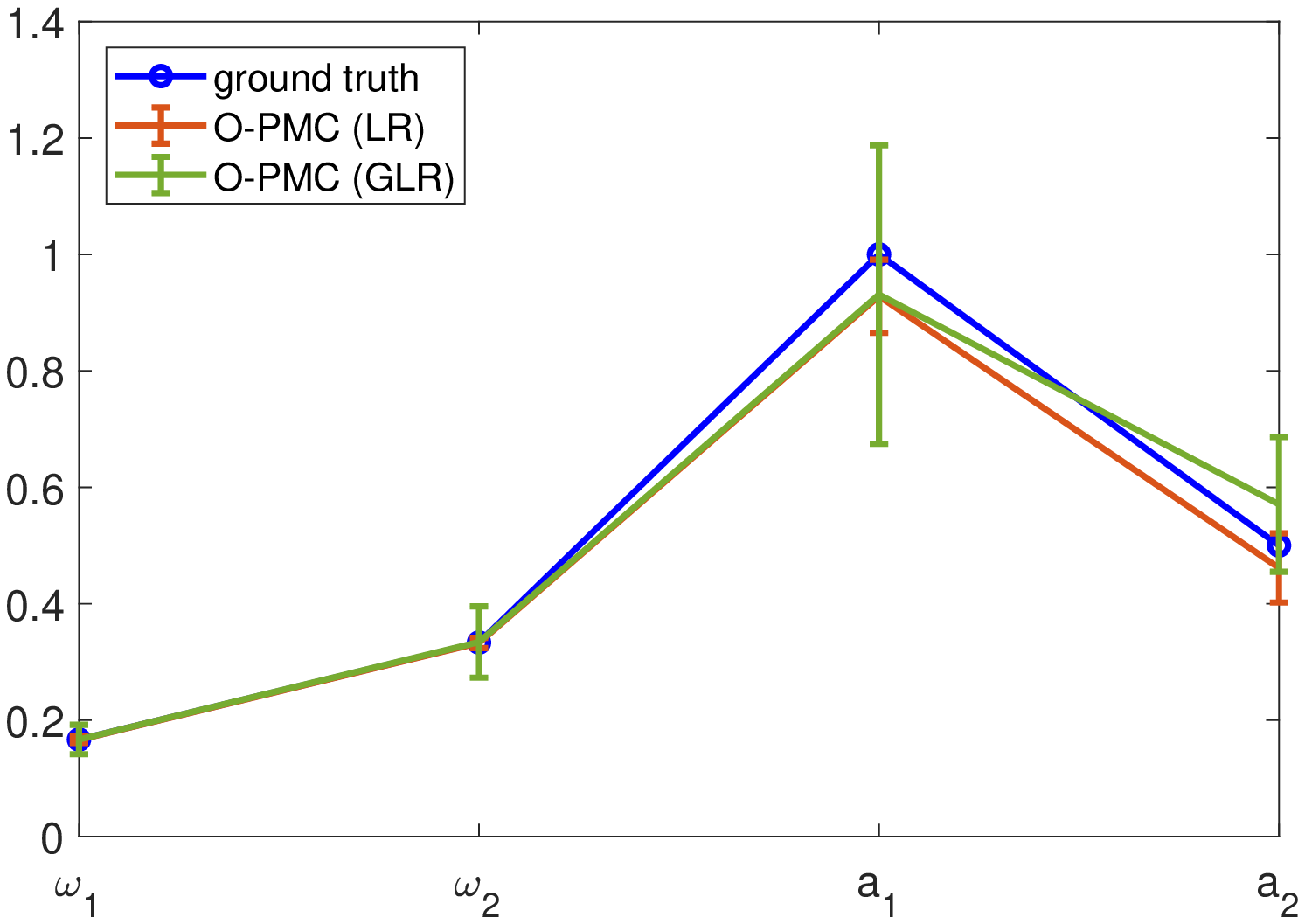}\\
\includegraphics[width=0.49\columnwidth]{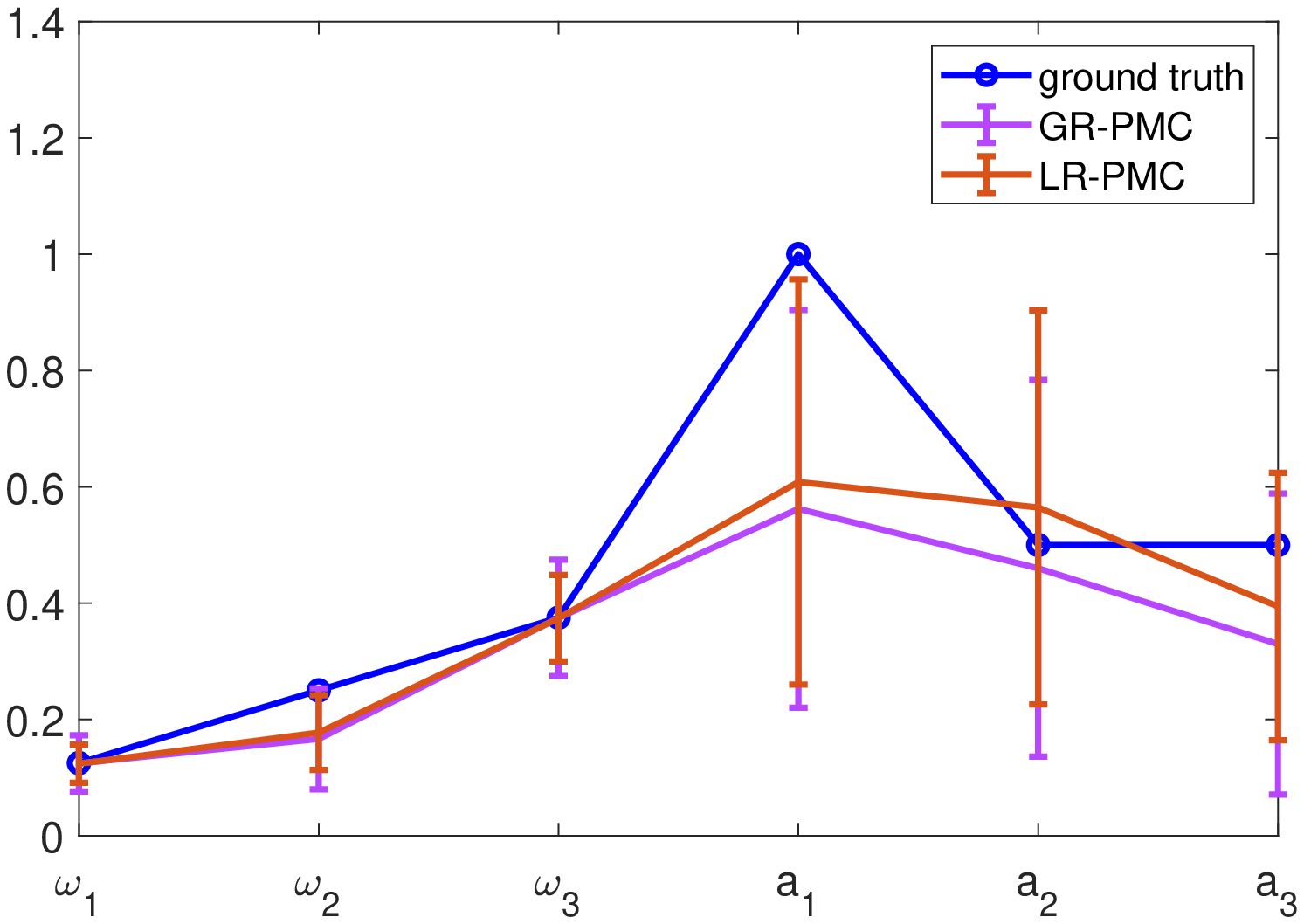} & \includegraphics[width=0.49\columnwidth]{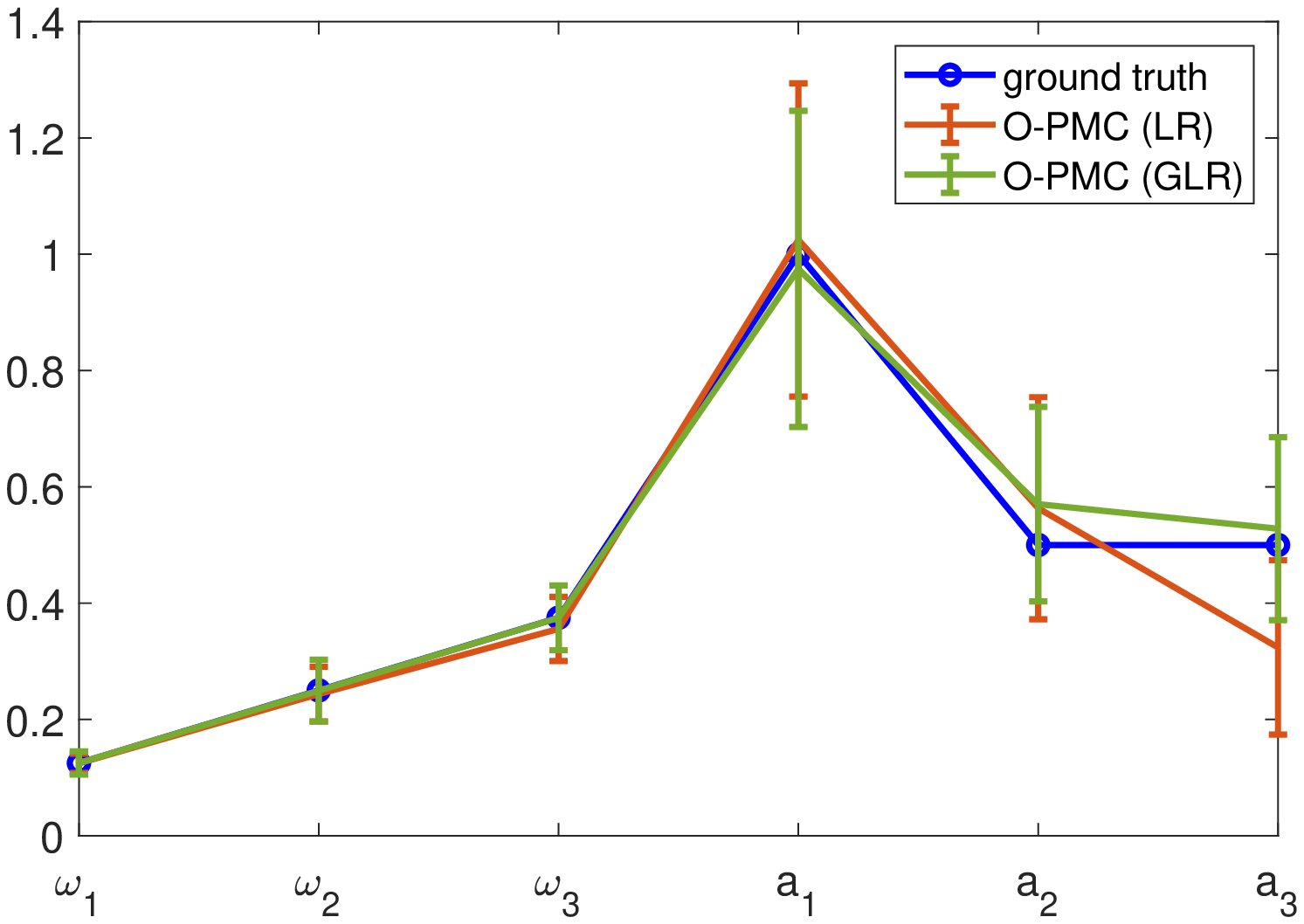}
\\
\includegraphics[width=0.49\columnwidth]{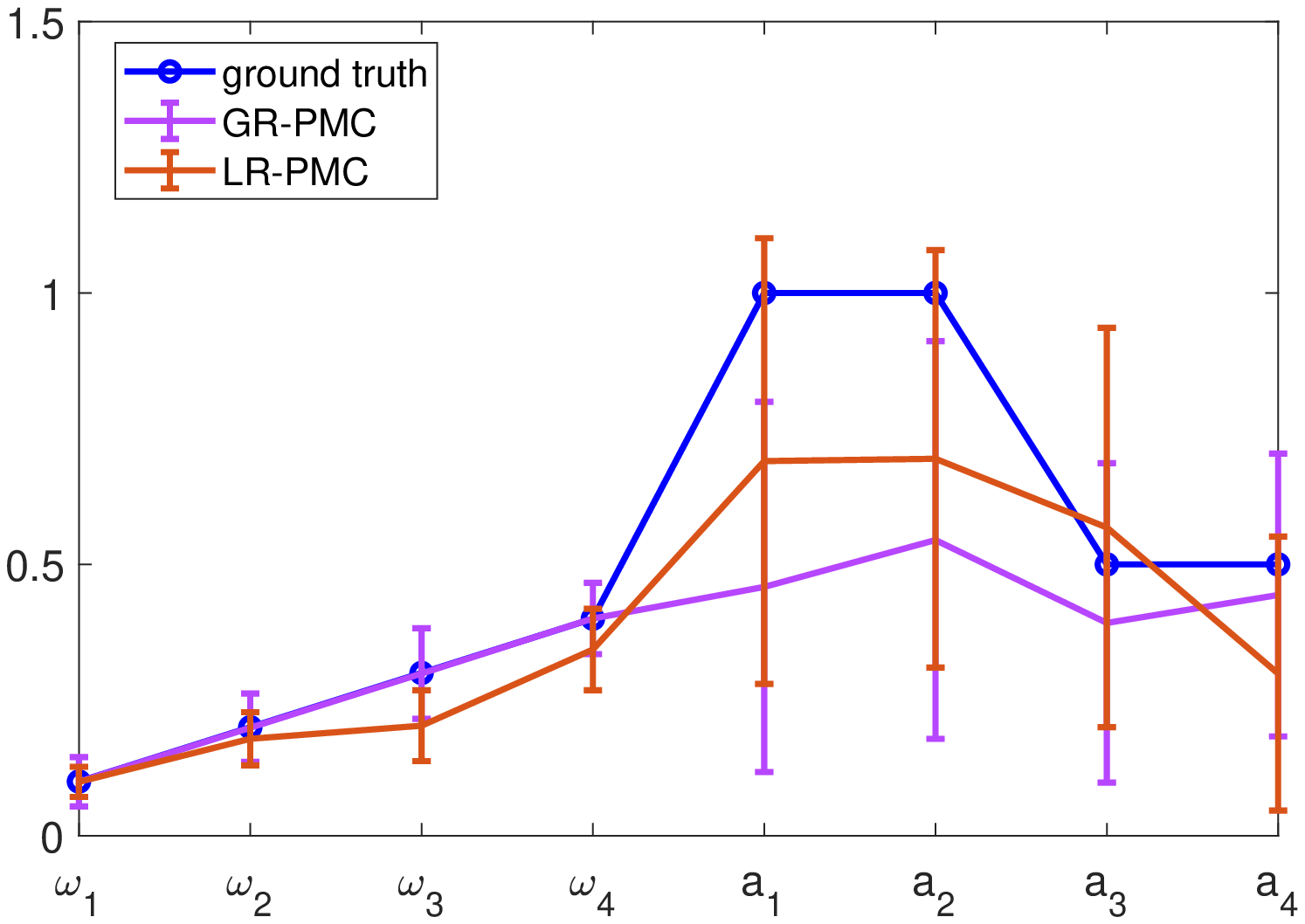} & \includegraphics[width=0.49\columnwidth]{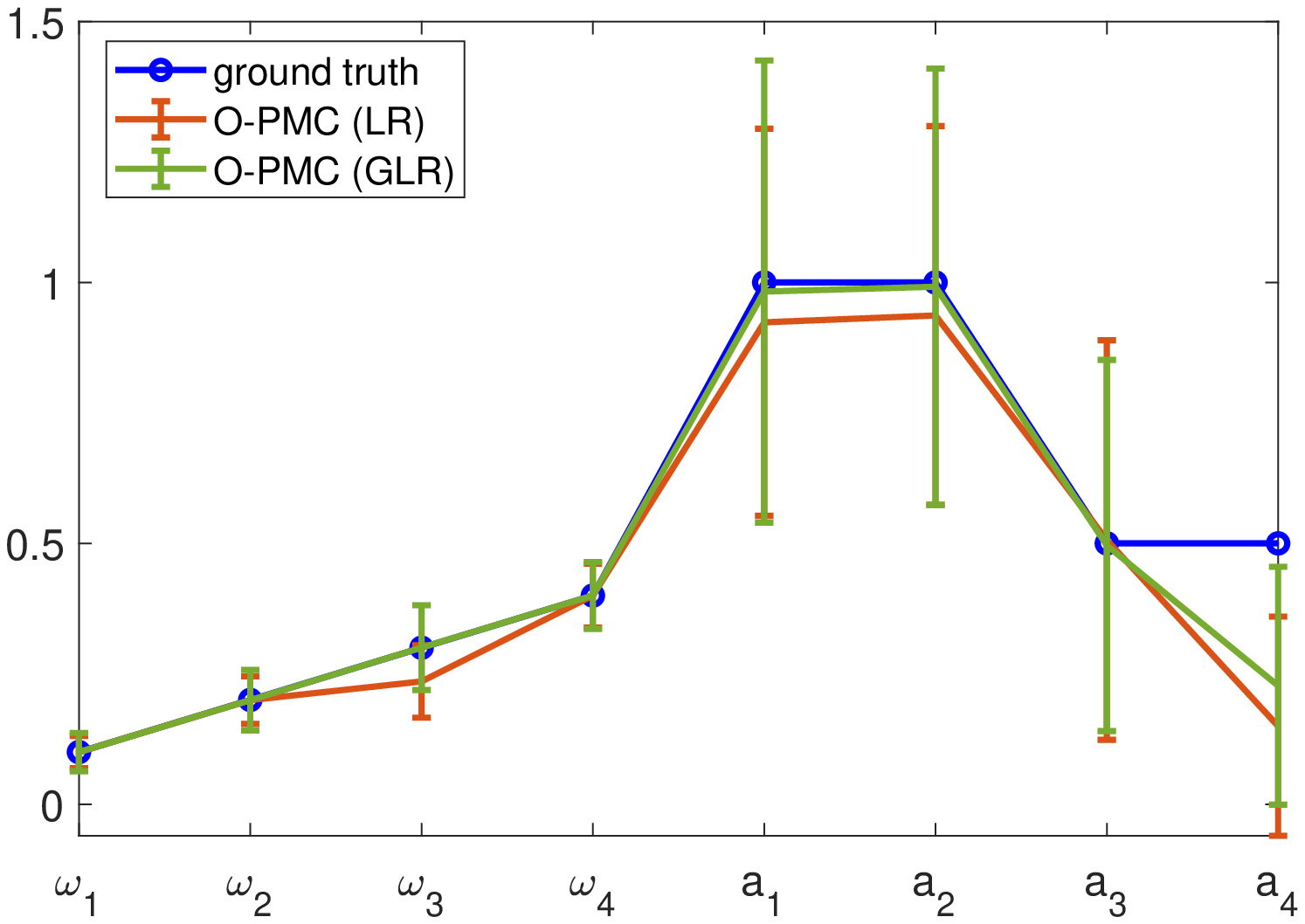}
\end{tabular}
\caption{\textbf{Example \ref{ex:sinus}.} Ground truth (blue) and estimated values {(median $\pm$ MAD of the mean estimator)} for frequencies and amplitudes in dimension $4$ (top), $6$ (middle) and $8$ (bottom), for GR-PMC, LR-PMC, and \acro \ using either LR or GLR scheme.} %$\emilie{particularly impressive in dimension 6.}}
%\label{fig:banana_slpmc}
\label{fig_freqs_comparison_gt}
\end{figure} 

\begin{table*} [h!]
\scriptsize{
\setlength{\tabcolsep}{2pt}
\def\marginwidth{1.5mm}
\begin{center}
\begin{tabular}{|c||c|c|c||c|c|c||c|c|c||c|c|c||c|c|}                                 
\hline 
&  \multicolumn{3}{c }{ GR-PMC} &  \multicolumn{3}{|c|}{ LR-PMC} &  \multicolumn{3}{|c|}{GAPIS}  & \multicolumn{3}{|c|}{AMIS} & \multicolumn{2}{||c|}{ \acro} \\
\cline{2-15}
%DM PMC 0.001 0.01 0.1
&  $\sigma=10^{-3}$  & $\sigma=10^{-2}$   & $\sigma=10^{-1}$  &  $\sigma=10^{-3}$  & $\sigma=10^{-2}$   & $\sigma=10^{-1}$  & $\sigma=10^{-3}$  & $\sigma=10^{-2}$   & $\sigma=10^{-1}$ & $\sigma=10^{-3}$  & $\sigma=10^{-2}$   & $\sigma=10^{-1}$ &  LR & GLR\\
\hline
\hline
$d_x = 4$ & 0.1083 &0.0479 &0.0249  &  0.0516 &0.0185 &0.0299&  0.8422 &0.4176 &0.3342 &   0.0623 &0.0384 &0.0504
&   \textbf{0.0017}  &  0.0024 \\ 
\hline
$d_x = 6$ &   0.0929 &0.0568 &0.0544 & 0.0808 &0.0598 &0.0621 &   3.9936 &4.9897 &4.0805 & 0.0881 &0.0956 &0.0806
&   0.0076  &  \textbf{0.0014} \\ 
\hline
$d_x = 8$ & 0.1163 &0.0906 &0.1022  & 0.1041 &0.0718 &0.1128&  13.0336 &10.4020 &7.3938  & 0.1837 &0.1459 &0.1261
&  0.0418   &  \textbf{0.0343} \\ 
\hline
$d_x = 10$ & 0.0671 &0.0804 &0.0671  & 0.0609 &0.0933 &0.0757&  18.7525 &18.5906 &14.8404  & 0.1279 &0.1284 &0.1811
&  0.1027  &  \textbf{0.0560}  \\ 
\hline                                                          
\end{tabular}
\end{center}
\caption{\textbf{Example \ref{ex:sinus}.} Median MSE with respect to ground truth amplitudes and frequencies parameters, for dimensions $d_x = 4$, $6$, $8$ and $10$. For \acro, we set the initial proposal variance to $\sigma = 10^{-2}$. The period for GLR is set to $\Delta = 5$. In all PMC-based methods, $(N,K,T) = (50,20,20)$ while $(N,K,T) = (1,500,40)$ for AMIS. %GLR with period 5
}
\label{table_exp3_median}
}
\end{table*}

\begin{table*} [h!]
\scriptsize{
\setlength{\tabcolsep}{2pt}
\def\marginwidth{1.5mm}
\begin{center}
\begin{tabular}{|c||c|c|c||c|c|c||c|c|c||c|c|c||c|c|}                                 
\hline 
&  \multicolumn{3}{c }{ GR-PMC} &  \multicolumn{3}{|c|}{ LR-PMC} &  \multicolumn{3}{|c|}{GAPIS}  & \multicolumn{3}{|c|}{AMIS} & \multicolumn{2}{||c|}{ \acro} \\
\cline{2-15}
%DM PMC 0.001 0.01 0.1
&  $\sigma=10^{-3}$  & $\sigma=10^{-2}$   & $\sigma=10^{-1}$  &  $\sigma=10^{-3}$  & $\sigma=10^{-2}$   & $\sigma=10^{-1}$  & $\sigma=10^{-3}$  & $\sigma=10^{-2}$   & $\sigma=10^{-1}$ & $\sigma=10^{-3}$  & $\sigma=10^{-2}$   & $\sigma=10^{-1}$ &  LR & GLR\\
\hline
\hline
$d_x = 4$ & 0.4206 &0.1715 & \textbf{0.1122}  &  0.3670 &0.2453 &0.5379 &  0.5538 &0.3475 &0.3810 &  0.2956 &0.2697 &0.3185
&   0.2081  &  0.3043 \\ 
\hline
$d_x = 6$ &   0.7301 &0.3572 &0.2677 & 0.6632 &0.3745 &0.4739&  0.8761 &0.6995 &0.6757 &  0.6480 &0.6631 &0.6811
&  0.3859   &  \textbf{0.1681}  \\ 
\hline
$d_x = 8$ &  1.3138 &0.6910 &0.8157  &  1.2692 &0.7238 &1.0867 &  1.5086 &1.2156 &1.5075  &  3.9607 &3.3690 &3.2748
&   0.5733  &  \textbf{0.4259} \\ 
\hline         
$d_x = 10$ &  2.6353 &1.0050 &2.8790  &  2.1628 &1.1288 &2.8382 & 1.4591 &1.5012 &1.5146&  4.5971 &4.5016 &4.7863
&   1.1124  & \textbf{0.7351}   \\ 
\hline                                                   
\end{tabular}
\end{center}
\caption{\textbf{Example \ref{ex:sinus}.} {Reconstructed MSE} for dimensions $d_x = 4$, $6$, $8$ and $10$. For \acro, we set the initial proposal variance to $\sigma = 10^{-2}$. The period for GLR is set to $\Delta = 5$. In all PMC-based methods, $(N,K,T) = (50,20,20)$ while $(N,K,T) = (1,500,40)$ for AMIS. %GLR with period 5
}
\label{table_exp3}
}
\end{table*}

\section{Conclusion}

{We have proposed the \acro \ algorithm, an AIS sampler of the family of PMC algorithms that incorporates geometric information of the target distribution. \acro \ exploits the benefits of the PMC framework, incorporates suitable resampling schemes, and includes efficient adaptive mechanisms. In particular, the novel algorithm adapts the location and scale parameters of a set of proposals. At each iteration, the location parameters are adapted through a suitable  resampling strategy combined with an advanced optimization-based scheme. The local second-order information of the target is exploited through a preconditioning matrix that acts as a scaling metric onto a gradient direction. This metric is also used in order to adapt the scale parameters of the proposals. We have discussed the choice of parameters, included an illustrative toy example, and evaluated numerically the performance of the novel algorithm in three challenging problems, comparing the results with state-of-the-art competitive methods. As a future work, we may explore the implementation of low-complexity approximations of the Hessian to adapt the scale parameters of the proposals.}

\bibliographystyle{ieeetr}

%\bibliography{bibliografia}
 
\end{document}